\documentclass[journal=jctcce,manuscript=article]{achemso}
\setkeys{acs}{usetitle = true}
\setkeys{acs}{maxauthors=10,etalmode=truncate}
\usepackage{xspace, amsmath, multirow, setspace, graphicx, subfigure}
\usepackage{bm, dcolumn, epsfig, color, xr, booktabs, longtable, lscape, titlecaps}
\mciteErrorOnUnknownfalse
\usepackage{soul}
\usepackage{xr}
\title[]{Reactive atomistic simulations of Diels-Alder-type reactions:
  Conformational and dynamic effects in the polar cycloaddition of
  2,3-dibromobutadiene radical ions with maleic anhydride}

\author{Ux\'ia Rivero} \affiliation{Department of
  Chemistry, University of Basel, Klingelbergstrasse 80, Basel,
  Switzerland}\altaffiliation{Contributed equally to this work}

\author{Haydar Taylan Turan} \affiliation{Department of
  Chemistry, University of Basel, Klingelbergstrasse 80, Basel,
  Switzerland}\altaffiliation{Contributed equally to this work}

\author{Markus Meuwly} \affiliation{Department of Chemistry,
  University of Basel, Klingelbergstrasse 80, Basel, Switzerland}
\email{m.meuwly@unibas.ch}

\author{Stefan Willitsch} \affiliation{Department of Chemistry,
  University of Basel, Klingelbergstrasse 80, Basel, Switzerland}
\email{stefan.willitsch@unibas.ch}


\begin{document}
%\doublespace

%%%%%%%%%% PRELIMINARY MATERIAL %%%%%%%%%%
\maketitle
\thispagestyle{empty}
%%%%%%%%%% MAIN TEXT STARTS HERE %%%%%%%%%%
\begin{abstract}
The kinetics, dynamics and conformational specificities for the ionic
Diels-Alder reaction (polar cycloaddition) of maleic anhydride with
2,3-dibromobutadiene radical ions have been studied theoretically
using multisurface adiabatic reactive molecular dynamics. A
competition of concerted and stepwise reaction pathways was found and
both the s-\textit{cis} and s-\textit{trans} conformers of the
diene are reactive. The analysis of the minimum dynamic
path of the reaction indicates that both, rotations and vibrations of
the reactant molecules are important for driving the system towards
the transition state. The rates were computed as $k = 5.1 \times
10^{-14}$ s$^{-1}$ for the s-\textit{cis} and $k = 3.8 \times
10^{-14}$ s$^{-1}$ for the s-\textit{trans} conformer of
2,3-dibromobutadiene at an internal temperature of 300~K. The present
results are to be contrasted with the neutral variant of the title
system in which only the \textit{gauche} conformer of the diene was
found to undergo a considerably slower, concerted and mostly
synchronous reaction driven by the excitation of rotations. The
results presented here inform detailed experimental studies of the
dynamics of polar cycloadditions under single-collision conditions in
the gas phase.
\end{abstract}
\maketitle

\section{Introduction}

\noindent
The Diels-Alder (DA) reaction in which a diene reacts with a
dienophile to form a cyclic product is a widely used tool in synthetic
chemistry \cite{diels28a, ishihara14a}. In this reaction, two $\rm
\sigma$ bonds and one $\rm \pi$ bond are formed from three $\rm \pi$
bonds with a high degree of regio- and stereoselectivity. Over the
past decades, a large number of experimental and theoretical studies
have been devoted to studying the mechanism of DA reactions and its
dependence on the geometric and electronic properties of the
reactants, see, e.g.,
Refs. \cite{houk95a,yepes13a,souza16a,domingo14a,
  horn96a,diau99a,saettel02a,singleton01a,goldstein96a,sakai00a,domingo09a,donoghue06a}
and references therein.\\

\noindent
Since two bonds are formed in this reaction, questions pertaining to
its concertedness and synchronicity are central to the understanding
of the reaction mechanism. A reaction is considered to be concerted if
the reaction pathway exhibits only a single transition state (TS)
between reactants and products so that it occurs in a single step. By
contrast, a stepwise mechanism involves several transition states
which have to be traversed between the reactants and the products. The
time elapsed between formation of the first and the second bond
defines the synchronicity of the process \cite{minkin99a}. A
synchronous process is necessarily concerted, but an asynchronous one
can be concerted or stepwise depending on the presence or absence of
intermediates. As an important implication, only the
s-\textit{cis}-conformer of the diene can react in a synchronous DA
reaction, whereas both the s-\textit{cis} and s-\textit{trans}
conformational isomers can in principle be reactive in a stepwise
mechanism.\\

\noindent
In the literature, there has been a long-standing discussion about the
synchronicity and concertedness of DA reactions \cite{houk95a,
  domingo14a}. The textbook picture of this reaction is that of a
concerted, synchronous process governed by the Woodward-Hoffmann rules
involving an aromatic TS \cite{hoffmann68a,ishihara14a}. However,
experiments and calculations have revealed many cases which deviate
from this paradigm. This is particularly the case for ionic DA
reactions (polar cycloadditions) in which one of the reagents is
oxidized to form a radical cation. Radical ionic variants of the DA
reaction are often faster than their neutral counterparts but still
show a high degree of stereoselectivity \cite{bellville82a, bauld87a,
  wiest92a}. A number of studies focused on the question of the
conservation of orbital symmetry in these ionic reactions in view of
the Woodward-Hoffmann rules which are widely used for rationalizing
mechanistic aspects of neutral DA processes \cite{wiest92a, bauld83a,
  chockalingam90a}.  In Ref. \cite{donoghue06a}, it was discussed that
arguments based on orbital symmetry can be misleading for polar
cycloadditions. In electronic-symmetry-conserving reactions, orbital concepts
should be replaced with an analysis of the symmetries of the
electronic states of the various species along the entire reaction
path. \\

\noindent
Experimentally, gas-phase collision studies carried out under
single-collision conditions represent a powerful tool to explore
dynamic effects in elementary chemical reactions.  In combination with
advanced product-detection techniques such as velocity-mapped ion
imaging (VMI) introduced by Eppink and Parker \cite{eppink97a}, they
enable the characterization of reaction mechanisms and dynamics in
unprecedented detail. In the specific context of DA reactions, previous
gas-phase experiments on the polar cycloaddition between butadiene
ions and ethene have been unable to isolate the DA product. As no
efficient deactivation of the cycloadduct was possible in the gas
phase, it was concluded that the product must have fragmented under
the experimental conditions \cite{buchoux94a}. This conclusion was
supported by subsequent computational studies which explored the
possible fragmentation pathways of the DA product in order to
interpret the experimental findings \cite{hofmann99a, bouchoux04a}. \\

\noindent
Here, the gas-phase ionic DA reaction between 2,3-dibromobutadiene
radical ions (DBB$^+$) and maleic anhydride (MA) was studied by means
of reactive molecular dynamics simulations. Previous work
\cite{rivero17a, rivero19b} had revealed that the neutral counterpart
of this reaction is synchronous, direct and promoted by rotational
excitation of the reactant molecules. By contrast, the present work
shows that the ionic system is characterized by a competition of
concerted and stepwise reaction pathways and that both, the
s-\textit{cis} and s-\textit{trans} conformers of DBB$^+$ are
reactive.  With computed rates of $k = 5.1 \times 10^{-14}$ s$^{-1}$
for the s-\textit{cis} and $k = 3.8 \times 10^{-14}$ s$^{-1}$ for the
s-\textit{trans} conformers of DBB$^+$, respectively, at an internal
temperature of 300~K and a collision energy of 100 kcal/mol, the ionic
reaction was found to be considerably faster than its neutral variant
under these conditions. Rotations still play a role in activating the
ionic reaction, but less pronouncedly than in the neutral system. The
present study highlights salient dynamic differences between neutral
and ionic DA reactions and paves the way for a detailed investigation
of these effects in conformationally controlled gas-phase experiments
\cite{willitsch17a}.\\

\section{Methods}

\begin{figure}
\includegraphics[width=0.95\textwidth]{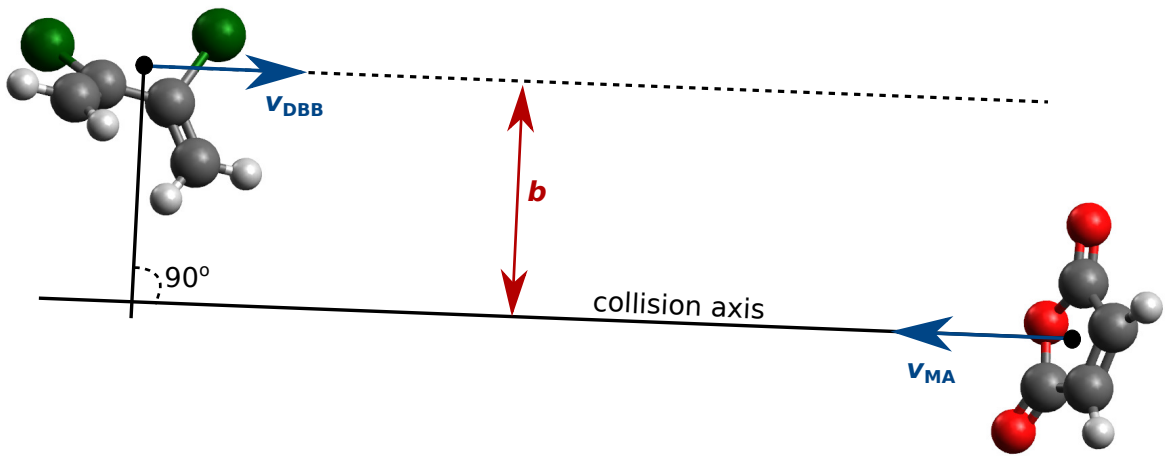}
\caption[Schematic of the initial conditions of a
  trajectory.]{Schematic of the initial conditions of a
  trajectory. The centers of mass of the reactant molecules were initially separated by 20
  \AA\/ along the collision axis. The impact parameter ($b$) was
  specified by displacing the 2,3-dibromobutadiene ion (DBB$^+$) along
   an axis perpendicular to the collision axis. The blue arrows represent the initial
  velocities of the centers of mass of DBB$^+$ ($v_{\rm DBB}$) and
  maleic anhydride (MA, $v_{\rm MA}$). }
\label{fig:figure1}
\end{figure}

\subsection{Molecular Dynamics Simulations}
Atomistic simulations were carried out with the CHARMM
program\cite{CHARMM} using multisurface adiabatic reactive molecular
dynamics (MS-ARMD).\cite{nagy14a} Initial conditions for the collision
simulations were generated from ensembles of the individual molecules
(MA and DBB$^+$) at different vibrational temperatures. Heating and
equilibration temperatures were selected according to the desired
final vibrational temperature ($T_{\rm vib}$). The centers of mass of
the two reactants were initially separated by 20 \AA\/ with a random
relative orientation of the molecules. The collision energy ($E_{\rm
  coll}$) was chosen by scaling the atomic velocities along the
collision axis. Rotational energy corresponding to a particular
rotational temperature ($T_{\rm rot}$) was added to the molecules
following calculation of their moment-of-inertia tensor and assuming
equipartition among the three rotational degrees of freedom
\cite{atkins94a}. The impact parameter ($b$) was uniformly sampled by
displacing the center of mass of one of the molecules along an axis
perpendicular to the collision axis (Figure \ref{fig:figure1}). Excitation of specific
vibrational modes was achieved by projecting the initial velocities
onto the space of normal modes and by modifying the kinetic energy of
the desired normal mode. All bonds, including those involving hydrogen
atoms, were kept flexible and the time step in the simulations was
sufficiently small ($\Delta t = 0.1$~fs) to ensure conservation of
total energy. For propagating the equations of motion, the velocity
Verlet algorithm was used.\cite{verlet1967computer}\\

\subsection{Force Field Parametrization}
For the reactive simulations MS-ARMD\cite{nagy14a} was used. Reference
calculations for the parametrization were carried out at the density
functional theory (DFT) level with the M06-2X
functional\cite{zhao2008m06} and the 6-31G* basis
set\cite{hehre1972self} using Gaussian09 for the electronic structure
calculations.\cite{g09} This level of theory was previously found to
yield an adequate description of the energetics of the system of
interest.\cite{rivero17a} For the initial force field of reactant and product states, the parameters from
SwissParam\cite{zoete11a} were used. Based on those, ensembles of
reactant- and product-state structures were generated with CHARMM as
follows: an optimization of the structures with the Newton-Raphson
method was followed by 50~ps of heating dynamics, 50~ps of
equilibration at 500~K, 60~ps of cooling down to 300~K and free
\textit{NVE} (microcanonical ensemble) dynamics. The temperature was
only raised up to 400~K for the
reactant van-der-Waals complex to avoid dissociation. For parametrising the intermediate, the final
temperature was set to 100~K to ensure obtaining low-energy
structures. Additional structures for this force field were generated
through scans around the first new bond formed along the reaction.\\

\noindent
Single point energies at the M06-2X/6-31G* level of theory were
computed for parametrizing the different force fields including the
product state (2086 structures), the intermediate (INT-tr$^+$, 1785
structures), the non-bonded interactions of the reactant (2589
structures), and the IRCs for the \textit{endo} (169 structures), the
\textit{exo} (192 structures) and the \textit{trans} (234 structures)
paths, respectively; see Section \ref{sec:param} below for a
discussion of these different structures. The harmonic bond, Morse
bond, angle and dihedral parameters of MS-ARMD force fields are
summarized in Tables S2 to S5 of the supplemental material (SM),
respectively. Further, non-bonded parameters of reactant, intermediate
and product PES are presented in Tables S6 to
S8 of the SM, respectively.  \\

\noindent
In the crossing region the force fields were connected by combining
the force fields of the reactants, intermediate and products with
``GAussian times POlynomial'' (GAPO) functions\cite{nagy14a}, see
supporting information. A genetic algorithm was used for fitting these
GAPOs.\cite{MM.hso3cl:2016} The global reactive potential energy
surface (PES) was thus
\begin{equation}
\label{ms-armd}
V_{\rm MS-ARMD}=\sum_{i=1}^{n} w_i(\textbf{x})
V_i(\textbf{x})+\sum_{i=1}^{n-1}
\sum_{j=1+1}^{n}[w_i(\textbf{x})+w_j(\textbf{x})] \sum_{k=1}^{n_{ij}}
\Delta V^{ij}_{\rm GAPO,k}(\textbf{x}),
\end{equation}
where $V_i(\textbf{x})$ is the energy of the force field of state $i$
(reactant, product, intermediate) at nuclear geometry $\textbf{x}$,
their weights $w_i(\textbf{x})$, and the $\Delta
V^{ij}_{\rm GAPO,k}(\textbf{x})$ are GAPO functions up to third (for
reactant and intermediate) and second (for intermediate and product)
polynomial order, respectively, see Table S9 of the SM. In order to
render the force field permutation invariant, two and four
different force fields for the description of the product and the
intermediate were used, respectively (see Tables S2 to
S8 of the SM).\\

\subsection{Analysis of the Trajectories}
Reactive trajectories were analyzed by decomposing the energy content
of the fragments along different degrees of freedom. For this purpose,
the total kinetic energy along the minimum dynamic
path\cite{unke2019b} (Section \ref{sec:mdp}) was analyzed in two
ways. In one approach, the total kinetic energy was projected onto the
eigenvectors of the Hessian matrix of the reactant molecules with
geometries corresponding to the last point of each
trajectory. Alternatively, the total kinetic energy was decomposed
into the translational energy of the center of mass of the reactant
molecules ($E_{\rm trans}$), and their rotational ($E_{\rm rot}$) and
vibrational ($E_{\rm vib}$) energy. The translational energies were
calculated according to
\begin{equation}
E_{{\rm trans}, \rm{A}} =\frac{\Big |\sum\limits_{i \in \rm{A}}
  \vec{p}_i \Big|^2}{2 M_{\rm A}},
\label{etrans}
\end{equation}
where $\vec{p}_i$ is the momentum of atom $i$ belonging to molecule A
(${\rm A} =$ MA, DBB$^+$) and $M_\text{A}$ is the total mass of
molecule A. The rotational energies were computed as
\begin{equation}
E_{{\rm rot}, \text{A}} =\frac{1}{2} | \textbf{I}_\text{A} \vec{\omega}_\text{A}^2|.
\label{erot2}
\end{equation}
Here, $\vec{\omega}_\text{A}$ is the angular velocity of molecule A and
$\mathbf{I}_\text{A}$ is the moment-of-inertia tensor of molecule A,
\begin{equation}
\vec{\omega}_\text{A} = \textbf{I}_\text{A}^{-1} \vec{L}_\text{A}.
\label{erot4}
\end{equation}	
In this equation, $\vec{L}_\text{A}$ is the angular momentum of
molecule $\text{A}$,
\begin{equation}
\vec{L}_\text{A} = \sum_{i \in \text{A}} \vec{r}_i~' \times \vec{p}_i~',
\label{erot3}
\end{equation}
where atomic momenta ($\vec{p_i}~'$) and atomic coordinates
($\vec{r_i}~'$) in the center of mass frame were calculated as:
\begin{equation}
{\vec{x}_i}~' = \vec{x}_i - \vec{x}_{{\rm CoM}, \text{A}};~~~~~ x = p,r
\label{erot1}
\end{equation}
and the subscript "${\rm CoM},\text{A}$" refers to the center of mass of
molecule A. Finally,
\begin{equation}
E_{{\rm vib}, \text{A}} = E_{{\rm tot}, \text{A}} - E_{{\rm rot}, \text{A}} - E_{{\rm
    trans}, \text{A}}
\label{evib}
\end{equation}	
where $ E_{{\rm tot}, \text{A}} $ is the total kinetic energy of molecule A
along the trajectory. \\

\noindent	
The trajectories were considered reactive and terminated when they
reached the product force field. The reactive cross section $\sigma$
was calculated according to
\begin{equation}
\label{impact-parameter}
\sigma = 2 \pi b_{\rm max} \frac{1}{N_{\rm tot}} \sum_{i=1}^{N_{\rm
    reac}} b_i,
\end{equation}
\noindent
where $b_{\rm max}$ is the maximum impact parameter (defined as the
impact parameter at which no reactions could be observed anymore),
$N_{\rm tot}$ is the total number of trajectories, $N_{\rm reac}$ is
the number of reactive trajectories and $b_i$ is the impact parameter
of the reactive trajectory $i$. \\

\section{Results and Discussion}
\subsection{Parametrization of the Reactive Force Fields}
\label{sec:param}

\begin{figure}
\includegraphics[width=0.75\textwidth]{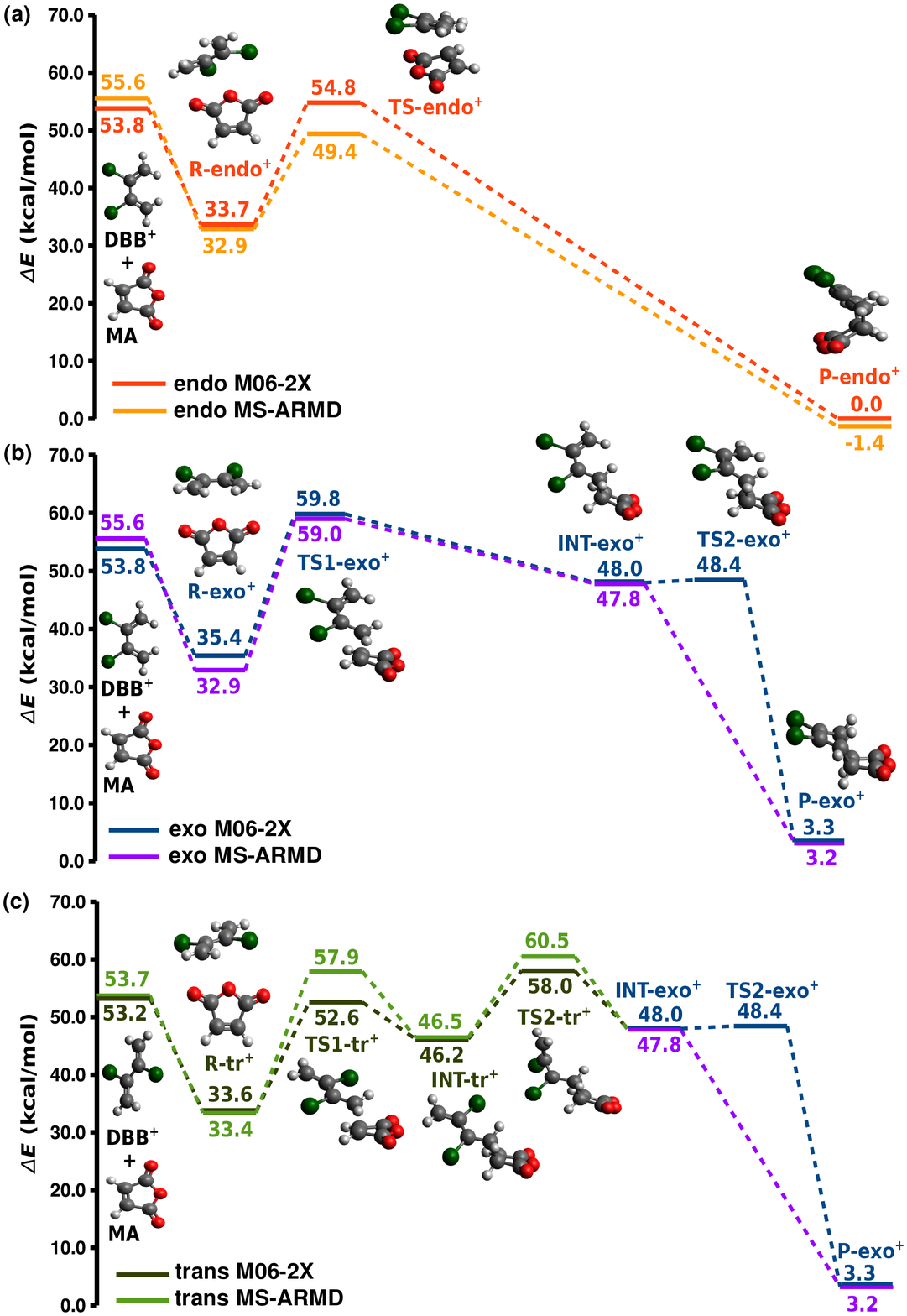}
\caption[Potential energy surface for the cationic Diels-Alder
  reaction between DBB$^+$ and MA]{Potential energy surface for the
  three possible DA reaction paths (a) \textit{endo}, (b)
  \textit{exo} and (c) \textit{trans} between 2,3-dibromobutadiene cation
  (DBB$^+$) and maleic anhydride (MA) at the M06-2X/6-31G* level of
  theory and from MS-ARMD. Relative energies are given in kcal/mol
  with respect to the \textit{endo} product (P-endo).The structures are connected by minimum-energy paths (indicated as dashed lines) verified by intrinsic-reaction-coordinate (IRC) calculations. The superscripts "$^+$" indicate ionic structures.}
\label{fig:figure2}  
\end{figure}
	
\begin{figure}
\includegraphics[width=0.65\textwidth]{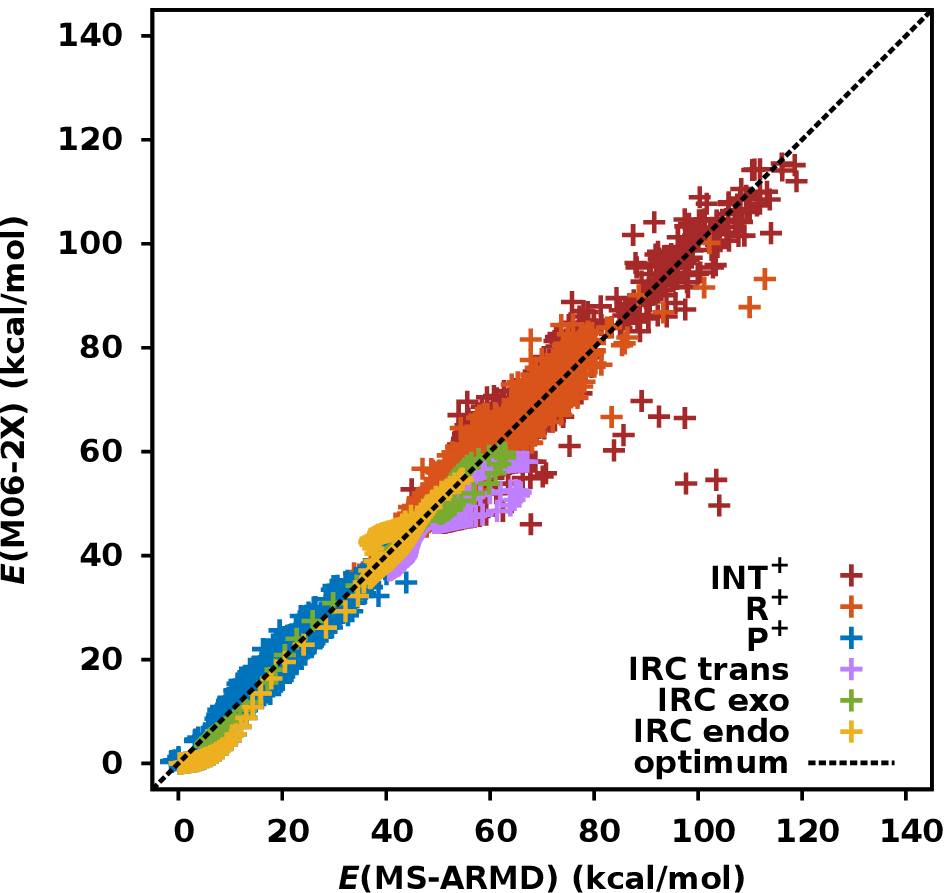}
\caption[Energy correlation of the MS-ARMD PES with the M06-2X/6-31G*
  level of theory]{\label{parametrizationcat}Energy correlation of
  7055 reference structures computed at the M06-2X/6-31G* level of
  theory and the MS-ARMD PES. The total root-mean-square deviation
  (RMSD) is 2.9~kcal/mol.}
\label{fig:figure3}
\end{figure}

\noindent
Figure \ref{fig:figure2} shows stationary points on the PES of the
Diels-Alder reaction between DBB$^+$ and MA at the M06-2X/6-31G* level
of theory \cite{rivero17a}. For the s-\textit{cis} conformer of
DBB$^+$ both reactant molecules (DBB$^+$ and MA) are symmetric. Thus,
there are two possible pathways for a concerted Diels-Alder reaction
referred to as ``\textit{endo}'' and ``\textit{exo}'' depending on the
relative orientation of the reactants (Figures \ref{fig:figure2} (a)
and (b)). For the \textit{exo} configuration, an additional stepwise
pathway via an intermediate INT-exo$^+$ was identified (Figure
\ref{fig:figure2} (b)). For the s-\textit{trans} conformer of DBB$^+$,
a stepwise pathway was found (\textit{"trans"}, Figure
\ref{fig:figure2} (c)). The \textit{endo} product (P-endo$^+$) was
defined as the zero of the energy scale in Figure
\ref{fig:figure2}. \\

\noindent
The aim of the present parametrization was to obtain a single,
globally valid reactive MS-ARMD PES that describes the three competing
paths, see Figure \ref{fig:figure2}, as had previously been done for
competitive ligand binding.\cite{MM.pathway:2013} The \textit{endo}
intrinsic reaction coordinate (IRC, Figure \ref{fig:figure2} (a)) was
used for parametrizing the GAPOs. It is important to mention that the
\textit{endo} IRC is asymmetric since it exhibits a plateau after the
transition state (see Figure S1 of the SM). The structures in this
region resemble those of the intermediate state with one of the new
C-C bonds formed. Hence, the intermediate force field is active in
this region which is an approximation because the \textit{endo} path
has no minimum there. However, this was the only viable way to
obtain a single global PES.\\

\noindent
The quality of the reactive PES compared with the reference DFT data
is reported in Figure \ref{fig:figure3}. The total root-mean-square
deviation (RMSD) is 2.9~kcal/mol over a range of 120~kcal/mol which is
deemed sufficient for a correct qualitative characterization of the
dynamics of the system. There are some outliers in the intermediate
force field (INT$^+$ in Figure \ref{fig:figure3}). However, because
they have high energies in the parametrized PES, the system will
rarely sample these regions which are thus expected to have a minor
impact on the dynamics. \\

\noindent
Figure \ref{fig:figure2} compares the MS-ARMD energies with those from
the reference DFT calculations. The \textit{exo} path (Figure
\ref{fig:figure2}(b)) is well described although TS2-exo$^+$ does not
exist on the MS-ARMD surface. As discussed above, the total RMSD of
the surface is 2.9~kcal/mol and so it is expected that a TS that lies
0.4 kcal/mol above the minimum INT-exo$^+$ is not captured by the
parametrized force field. The \textit{endo} path (Figure
\ref{fig:figure2}(a)) exhibits a TS 5.4~kcal/mol lower than the
reference energy which implies that the Diels-Alder reaction along
this path is more favorable when treated with the MS-ARMD PES than the
dissociation of the van der Waals complex in the entrance channel,
while at the DFT level the heights of the barriers towards
dissociation and the onward reaction are similar. This mismatch will
lead to overestimating the reaction rate along the \textit{endo} path
in MS-ARMD. Finally, for the \textit{trans} path (Figure
\ref{fig:figure2}(c)), the energies of TS1-tr$^+$ and TS2-tr$^+$ are
overestimated by 5.3~kcal/mol and 2.5~kcal/mol, respectively. The
overestimation of the energy of the first TS should not be worrisome,
because the bottleneck for the reaction along this path is TS2-tr$^+$
which lies higher in energy than TS1-tr$^+$ for both, the MS-ARMD PES
and the reference DFT calculations. However, the higher energy of
TS1-tr$^+$ in the MS-ARMD treatment will artificially extend the
lifetime of INT-tr$^+$. \\

\noindent
In addition to comparing energies for stable and transition states,
their geometries and harmonic frequencies were determined from the
MS-ARMD PES and from the reference DFT calculations, see Figures
S2 and S3 and Table S1 of the SM. A
superposition of the reactant, intermediate, transition-state and
product structures is shown in Figure S2 of the SM and the root mean
squared differences are reported in Table S1 of the SM. The monomeric
structures superimpose to within better than 0.1 \AA\/ which indicates
that the bonded parameters of the MS-ARMD force field are
reliable. For the complex structures the product and intermediate
states show deviations of up to 0.2 \AA\/ which increase to $\sim 0.3$
\AA\/ for transition state structures. This suggests that further
optimization of the nonbonded parameters (charges and van der Waals)
may be possible. The harmonic frequencies along the \textit{endo} path
from MS-ARMD and the DFT calculations agree very favorably, see
Figure S3a of the SM which underlines the quality of the reaction
path for which the GAPOs were parametrized. These parameters do,
however, not yield the same quality for the \textit{exo} and
\textit{trans} paths, in particular for the intermediate and high
frequencies for the INT$^+$ and TS2$^+$ structures (Figures S3b and
c). Of course, dedicated parametrization of these two paths with
increased accuracy would be possible but only at the expense of a
reduced generality of the global energy function.\\

\noindent
It is important to remember that the intermediate region of the PES is
quite flat reflecting that the intermediate structure is much more
flexible and samples a wider range of conformations compared to the
reactants or the products. Therefore, the reference data points on
this part of the surface are expected to be of lower quality than
those in the reactant and the product regions \cite{rivero17a}. For
this reason, special care is needed in the analysis of trajectories
that extensively sample the intermediate-state region of the PES where
the parametrization is less accurate than for the reactant and product
geometries. Because the aim was to develop a single reactive force
field for the s-\textit{cis} intermediate along the \textit{exo} path
(Figure \ref{fig:figure2}(b)) and the \textit{trans} intermediate
(Figure \ref{fig:figure2}(c)) in order to arrive at a global treatment
of all reaction pathways, the charges and equilibrium distances of
bonds and angles are identical for both these intermediates.

\subsection{Minimum Dynamic Path}
\label{sec:mdp} 

\begin{figure}[H]
  \includegraphics[width=0.9\textwidth]{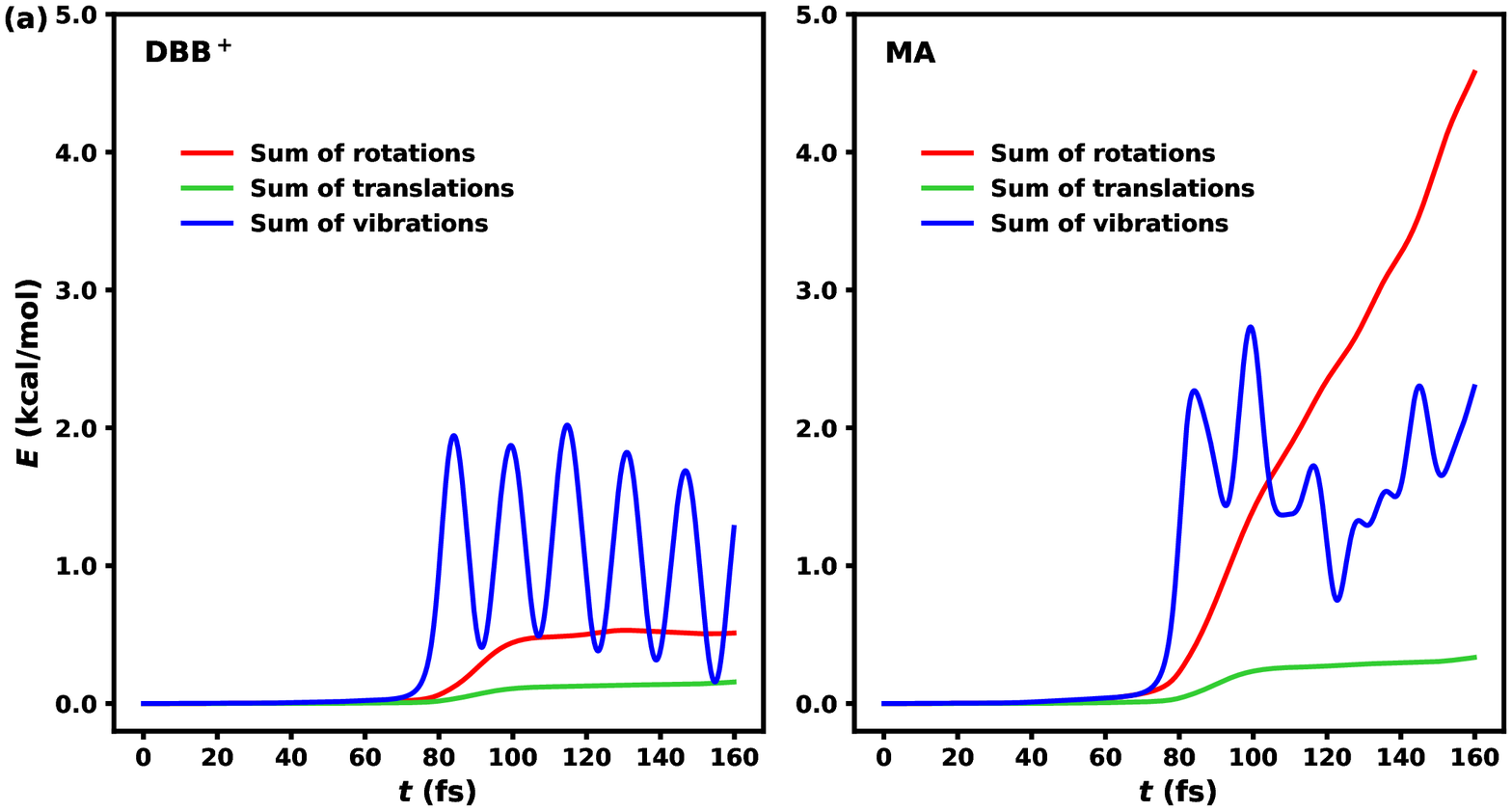}\\
  \vskip-10pt \includegraphics[width=0.9\textwidth]{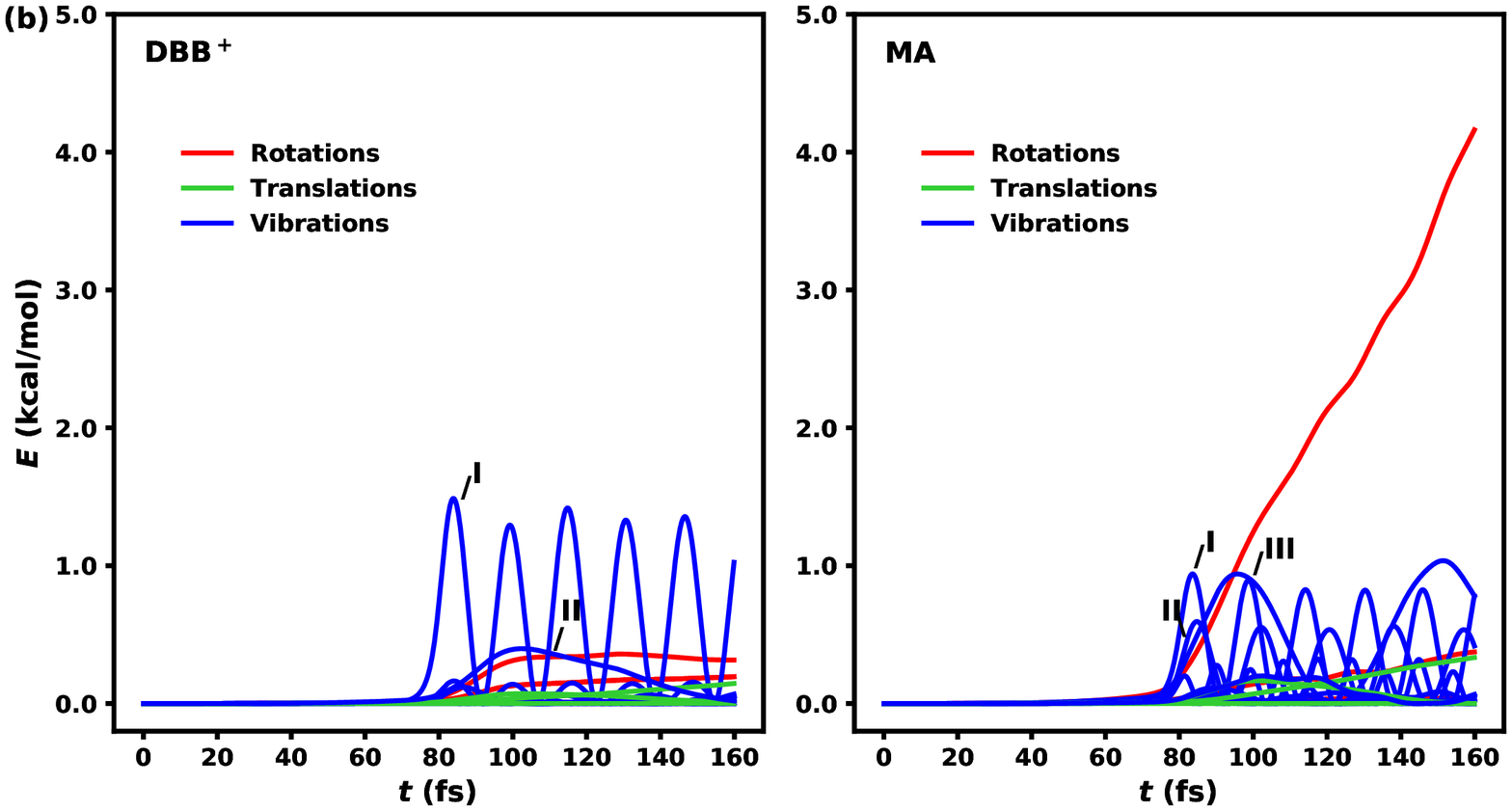}
  \caption{The MDP starting from TS1-exo$^+$. Projection of the total
    kinetic energy ($E$) onto the degrees of freedom of the
    2,3-dibromobutadiene ion (DBB$^+$) and maleic anhydride (MA) along the
    minimum dynamic path for the reaction of
    s-\textit{cis}-DBB$^+$~+~MA (a) summed into rotations,
    translations and vibrations and (b) further decomposed into
    individual components of the different degrees of freedom. The
    predominant active vibrations identified for DBB$^+$ are: (I)
    out-of-plane symmetric bending of hydrogens, and (II) skeleton
    out-of-plane asymmetric bend (\textit{cis/trans} isomerization
    mode); for MA: (I) and (II) asymmetric and symmetric out-of-plane
    hydrogen bending, respectively, and (III) asymmetric C=C
    out-of-plane bending.}
\label{fig:figure4}
\end{figure}

\begin{figure}[H] %[htb]
\includegraphics[width=\textwidth]{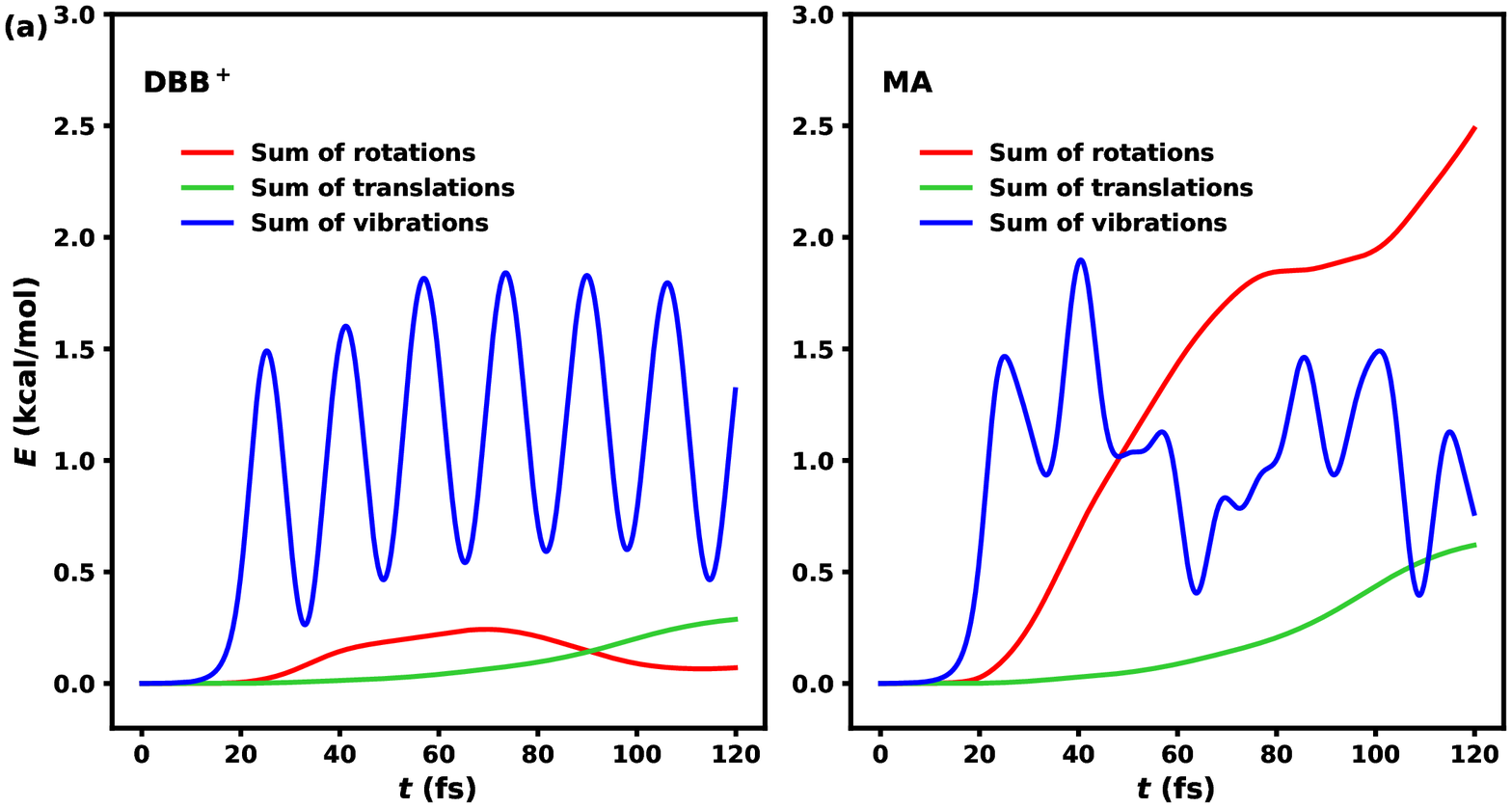}\\\vskip-10pt
\includegraphics[width=\textwidth]{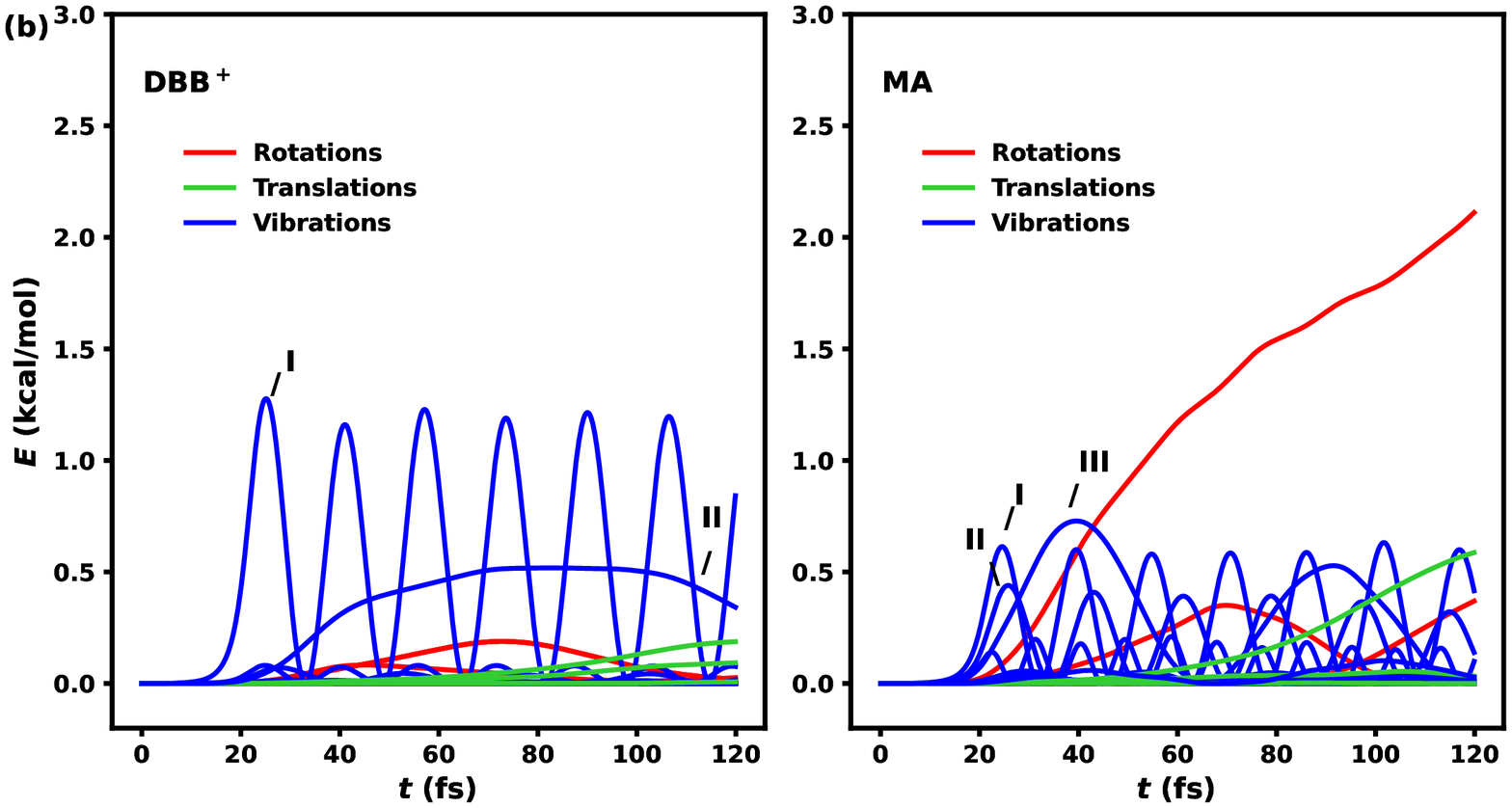}
\caption{The MDP starting from TS-endo$^+$. Projection of the total
  kinetic energy ($E$) onto the degrees of freedom of the
  2,3-dibromobutadiene ion (DBB$^+$) and maleic anhydride (MA) along the
  minimum dynamic path for the reaction of s-\textit{cis}-DBB$^+$~+~MA
  (a) summed into rotations, translations and vibrations (b) as
  individual traces. The predominant active vibrations identified for
  DBB$^+$ are: (I) out-of-plane symmetric bending of hydrogens and
  (II) skeleton out-of-plane asymmetric bend (\textit{cis/trans}
  isomerization mode); for MA: (I) and (II) asymmetric and symmetric
  out-of-plane hydrogen bending, respectively and (III) asymmetric C=C
  out-of-plane bending.}
\label{fig:figure5}
\end{figure}

\begin{figure}[H] %[htb]
\includegraphics[width=\textwidth]{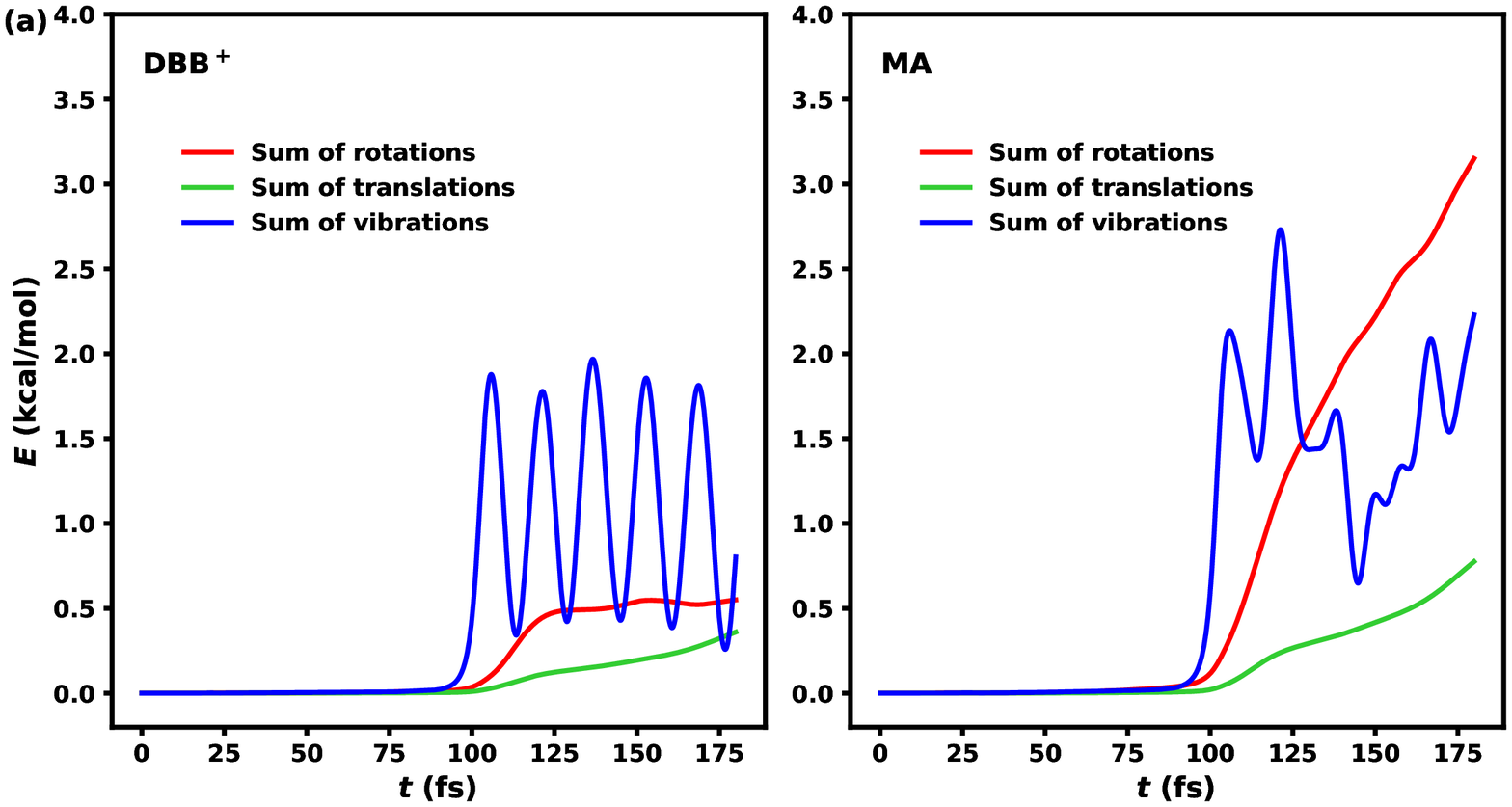}\\\vskip-10pt
\includegraphics[width=\textwidth]{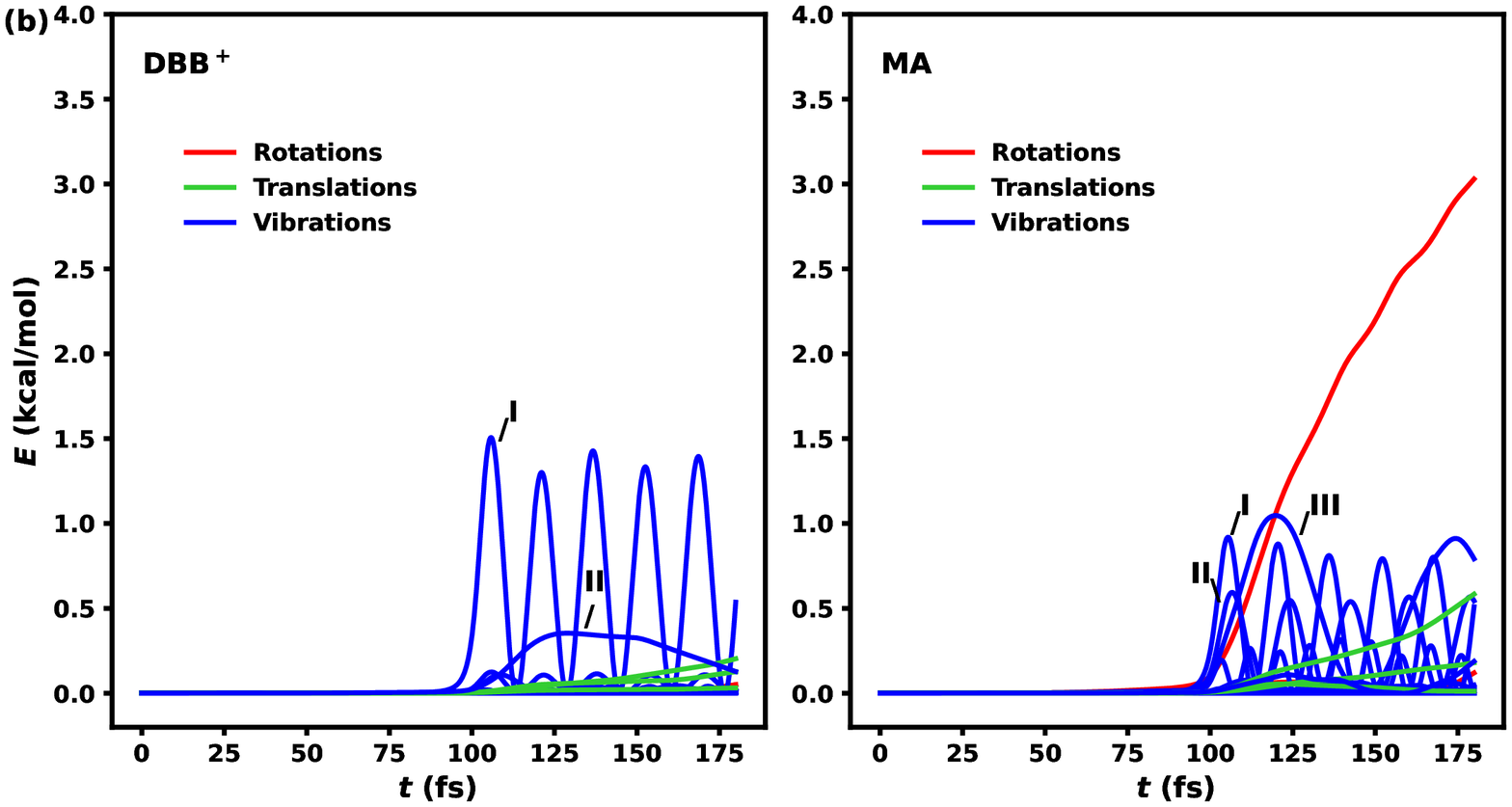}
\caption{The MDP starting from TS1-trans$^+$. Projection of the total
  kinetic energy ($E$) onto the degrees of freedom of the
  2,3-dibromobutadiene ion (DBB$^+$) and maleic anhydride (MA) along the
  minimum dynamic path for the reaction of
  s-\textit{trans}-DBB$^+$~+~MA (a) summed into rotations,
  translations and vibrations (b) as individual traces. The
  predominant active vibrations identified for DBB$^+$ are: (I)
  out-of-plane symmetric bending of hydrogens and (II) skeleton
  out-of-plane symmetric bend; for MA: (I) and (II) asymmetric and
  symmetric out-of-plane hydrogen bending, respectively and (III)
  asymmetric C=C out-of-plane bending.}
\label{fig:figure6}
\end{figure}

\noindent	
The minimum dynamic path (MDP) is the lowest-energy dynamical path
that follows Newton's equations of motion in phase
space.\cite{unke2019b} A trajectory starting at a TS geometry without
kinetic energy follows the MDP. The MDP was calculated for the three
different reaction pathways in the same fashion as previously for the
neutral reaction in Ref. \cite{rivero19b}. The following discussion
will be centered around the \textit{exo} path because it is best
described by the present MS-ARMD PES (Figure
\ref{fig:figure4}). However, since the \textit{endo} and
\textit{trans} paths are energetically more favorable, the MDPs of
these two pathways are also shown (Figures \ref{fig:figure5} and
\ref{fig:figure6}).\\

\noindent
For the \textit{exo} pathway, the projection of the total kinetic
energy along the MDP towards the reactants onto the degrees of freedom
of DBB$^+$ and MA is shown in Figure \ref{fig:figure4}(a) as sums of
the translational, rotational and vibrational energies. At $t=0$~fs
the system is at TS1-exo$^+$ and at $t=160$~fs it has arrived at the
reactants state. By projecting the total kinetic energy onto the
different degrees of freedom of DBB$^+$ and MA, the active degrees of
freedom in this reaction could be identified. Figure
\ref{fig:figure4}(a) shows that the largest amount of energy is
partitioned into the vibrations of DBB$^+$, while rotations contain
the largest amount of energy for MA although vibrations are also
active (see individual contributions in Figure
\ref{fig:figure4}(b)). The rotational energy of DBB$^+$ and MA
together accounts for 46\% of the total kinetic energy while
vibrational energy accounts for 48\% and translational energy for only
6\%. The same result was obtained from the direct decomposition of the
total kinetic energy (see Figure S4 of the SM). This finding stands
in clear contrast to the neutral DBB + MA system explored in
Ref. \cite{rivero19b} for which rotations accounted for 63\% of the total
kinetic energy and vibrations and translations for only 19\% and 18\%,
respectively.\\

\noindent		
The pronounced excitation of vibrational modes can be traced back to
the asymmetry of the cationic TS the breakup of which deforms the
molecules more strongly than in the neutral variant of the reaction
which exhibits a symmetric TS. On the grounds of microscopic
reversibility, the excitation of these vibrations is expected to
promote the reaction on its way towards the TS which provides valuable
information about future, possible experiments on this system.\\

\noindent
The projection of the total kinetic energy onto the degrees of freedom
of DBB$^+$ and MA along the \textit{endo} and \textit{trans} MDPs are
reported in Figures \ref{fig:figure5} and \ref{fig:figure6},
respectively. For the \textit{endo} path, vibrational energy accounts
for 54\% of the total kinetic energy while rotational and
translational degrees of freedom contain 38\% and 8\%,
respectively. This suggests that rotational energy is less important
to drive the reaction along the \textit{endo} path compared to the
\textit{exo} path, see Figure \ref{fig:figure4}. For the
\textit{trans} pathway, the contributions of vibrational, rotational
and translational energy are 50\%, 40\% and 10\%, respectively. The
active vibrations of MA were found to be the same in all paths. For
DBB$^+$, they remain the same for the \textit{endo} and \textit{exo}
trajectories and one mode changes for the \textit{trans} path because
the conformation of the molecule is different.\\

\noindent	
The MDP for the \textit{cis/trans} isomerization of INT$^+$ has also
been calculated. The total kinetic energy along this trajectory has
been projected onto the degrees of freedom of INT-tr$^+$ as shown in
Figure S5 of the SM. The energy is
essentially exclusively partitioned into vibrations as is expected for
a unimolecular reaction. The most active vibration is the
\textit{cis/trans} isomerization mode. Other low frequency skeleton
vibrations are also slightly active. \\

\subsection{Cross sections for the formation of van-der-Waals complexes in the entrance channel}

\begin{figure}[H]
\includegraphics[width=0.8\textwidth]{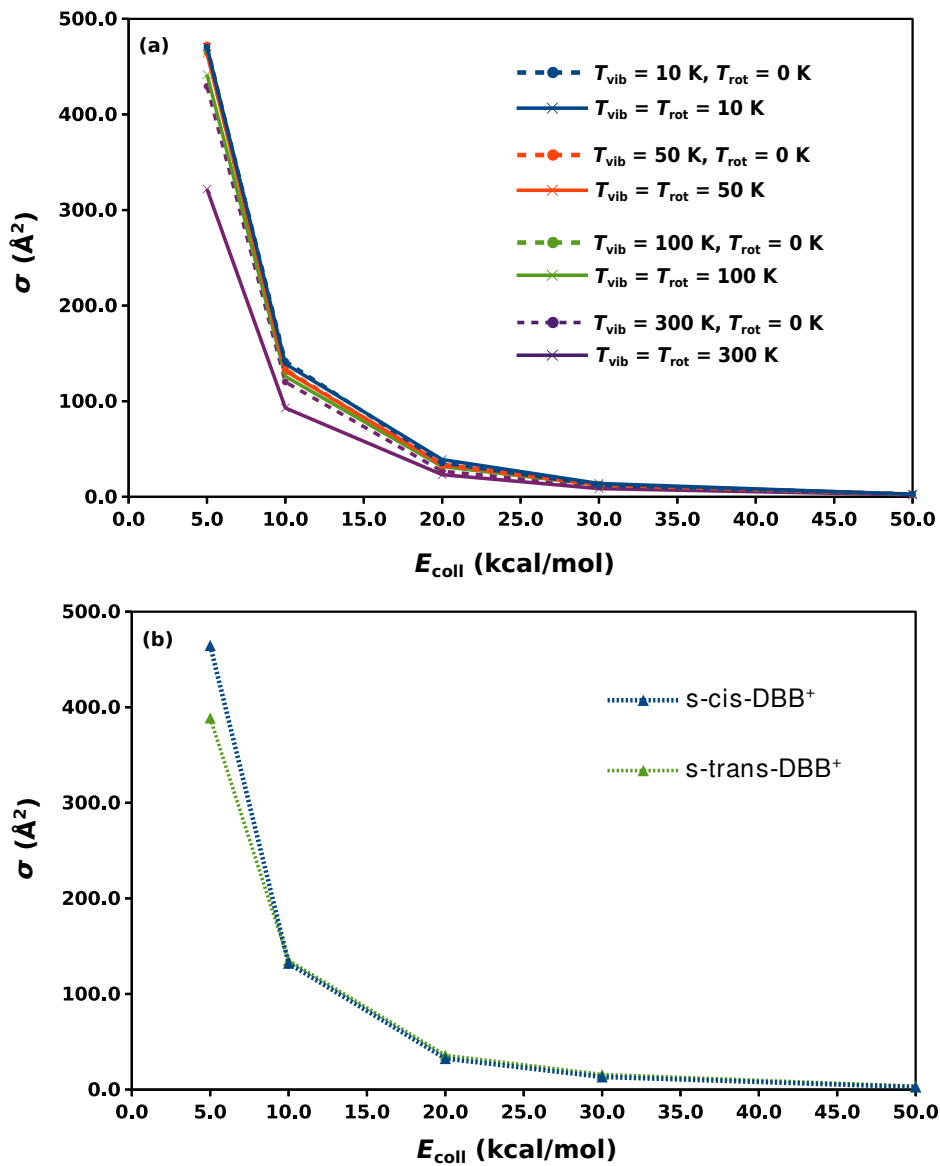}
\caption{(a) Variation of the cross section ($\sigma$) for the
  formation of the van-der-Waals complex in the entrance channel of
  the Diels-Alder reaction between 
  s-\textit{cis}-2,3-dibromobutadiene ions (DBB$^+$) and maleic anhydride (MA)
  as a function of the collision energy ($E_{\rm coll}$) at different
  vibrational and rotational temperatures ($T_{\rm vib}$, $T_{\rm
    rot}$). (b) Comparison of the cross sections for the
  s-\textit{cis} and s-\textit{trans} conformers of DBB$^+$ at $T_{\rm
    vib}=100$~K and $T_{\rm rot}= 0$~K.}
\label{fig:figure7}
\end{figure}

\noindent
The formation of van-der-Waals complexes in the entrance channel was
studied in order to establish whether the reaction is direct (i.e.,
without the formation of complexes) or complex-mediated. The impact
parameter $b$ was uniformly sampled in intervals of 1 \AA\/ up to a
maximum value $b_{\rm max}$ at which no reactive collisions could be
observed anymore. For each set of initial conditions ($E_{\rm coll}$,
$T_{\rm vib}$, $T_{\rm rot}$, $b$), 500~trajectories were run for
10~ps. If at the end of a trajectory the center-of-mass distance
between the two molecules was $<15$ \AA\/, it was
concluded that a van-der-Waals complex had been formed. Figure
\ref{fig:figure7}(a) shows the cross section $\sigma$ for the
formation of complexes as a function of the collision energy. It can
be seen that $\sigma$ diminishes as the collision energy
increases. Comparing to the neutral variant of the reaction studied in
Ref. \cite{rivero19b}, the maximum cross section computed is
$\sigma_\text{max} \approx 475~\rm \AA^2$, while for the neutral case
$\sigma\approx 110~\rm \AA^2$ was found. In addition, the cross
section for the ionic reaction decreases much slower with collision
energy and only totally vanishes at $E_{\rm coll} >
50$~kcal/mol. Rotational and vibrational energy have less influence
than in the neutral case. All these results reflect the fact that the
van-der-Waals complex in the entrance channel of the ionic reaction is
$\approx 12$~kcal/mol more stable than the one in the neutral
variant. \\

\noindent	
The influence of the initial conformation of DBB$^+$ in the cross
section for the formation of a van der Waals complex in the entrance
channel is shown in Figure \ref{fig:figure7}(b). It can be seen that
there is only a small difference at the lowest collision energies at
which the cross section for the s-\textit{cis} species is higher than
for the s-\textit{trans} conformer due to the fact that the maximum
impact parameter for complex formation was found to be $b_{max} = 16$
\AA\/ for s-\textit{cis}-DBB$^+$ while it is 14 \AA\/ for
s-\textit{trans}-DBB$^+$. However, these differences are judged to be
too small to be really significant.\\

\subsection{Reaction cross sections and rates}
The dynamics of the full reaction was investigated in two
steps. First, head-on collisions (i.e., $b=0$) were considered in
order to obtain an overview of the number of trajectories required and
the reaction rates to be expected. This was followed by a more
comprehensive study of off-axis collisions with $b>0$. Such a
procedure is warranted based on the previous finding for the neutral
reaction in which $\sim 10^7$ trajectories only lead to $\sim 500$
reactive events.\cite{rivero19b} Also, head-on collisions were found
to be most effective for the neutral reaction.\\

\begin{table}[H]
\caption{Initial conditions sampled for the recorded reactive events
  in terms of collision energy ($E_{\rm coll}$), conformation of the
  2,3-dibromobutadiene ion (DBB$^+$) and rotational temperature ($T_{\rm
    rot}$) at a vibrational temperature $T_{\rm vib} = 100$~K and
  impact parameter $b=0$~\AA. All simulations were propagated until
  dissociation or until they reached the products up to a total time
  of $t=600$~ps except for those with initial conditions $E_{\rm coll}
  = 50$~kcal/mol, $T_{\rm rot}=0$~K that were only propagated until $t
  =300$~ps.}
 \label{table-cat}
    
\resizebox{\textwidth}{!}{
\begin{tabular}{c|c|c|c|c|c}
  $E_{\rm coll}$ (kcal/mol) & DBB$^+$ & $T_{\rm rot}$ (K) & \# Products & \# Intermediates & \# Complexes \\  \hline
     & & & & & \\[-1.5ex]
  	50* & s-\textit{cis} & 0 & 1 & 19 & 13644 \\[0.5ex]
 	50* & s-\textit{trans} & 0 & 0 & 7 & 14828 \\[0.5ex]
 	50 & s-\textit{cis} & 2000 & 24 & 14 & 829 \\[0.5ex]
 	50 & s-\textit{trans} & 2000 & 7 & 25 & 884 \\[0.5ex]
 	50 & s-\textit{cis} & 4000 & 57 & 34 & 33 \\[0.5ex]
 	50 & s-\textit{trans} & 4000 & 29 & 40 & 50 \\[0.5ex]
  	75 & s-\textit{cis} & 0 & 18 & 13 & 104 \\[0.5ex]
 	75 & s-\textit{trans} & 0 & 4 & 3 & 174 \\[0.5ex]
 	75 & s-\textit{cis} & 2000 & 43 & 20 & 5 \\[0.5ex]
 	75 & s-\textit{trans} & 2000 & 22 & 15 & 17 \\[0.5ex]
 	75 & s-\textit{cis} & 4000 & 106 & 19 & 5 \\[0.5ex]
 	75 & s-\textit{trans} & 4000 & 69 & 29 & 23 \\[0.5ex]
  	100 & s-\textit{cis} & 0 & 30 & 4 & 0 \\[0.5ex]
 	100 & s-\textit{trans} & 0 & 6 & 0 & 0 \\[0.5ex]
 	100 & s-\textit{cis} & 2000 & 92 & 0 & 0 \\[0.5ex]
 	100 & s-\textit{trans} & 2000 & 47 & 3 & 2 \\[0.5ex]
 	100 & s-\textit{cis} & 4000 & 153 & 0 & 0 \\[0.5ex]
 	100 & s-\textit{trans} & 4000 & 65 & 0 & 0 \\ [0.5ex]\hline
 	& \multicolumn{2}{c|}{} & & & \\ [-1.5ex]
   Total &\multicolumn{2}{c|}{$1.8\cdot 10^6$ trajectories} & 773 & 245 & 30598
  \end{tabular} }
  \end{table}
  
\noindent  
For studying the full reaction, $1.8\cdot10^6$ MD simulations were
carried out. The vibrational temperature was set to
$T_\text{vib}=100$~K to mimic
vibrationally cold molecules in collision
experiments involving supersonic molecular beams and trapped ions
\cite{chang13a,kilaj18a}. Collision energies of $E_\text{coll}=50, 75$
and 100~kcal/mol were sampled. The rotational temperatures considered
were $T_\text{rot}=0, 2000$ and 4000~K such that the influence of
rotational excitation could be studied. The trajectories started with
either s-\textit{cis}-DBB$^+$ or s-\textit{trans}-DBB$^+$ (see
Table~\ref{table-cat}). However, trajectories starting in the
s-\textit{trans} conformer need not necessarily follow the
\textit{trans} path (Figure \ref{fig:figure2} (c)) as DBB$^+$ can
isomerize upon collision with MA. \\

\noindent	
The trajectories were propagated until a) dissociation of the
van-der-Waals complex back to the products occurred, b) the products
were formed, or c) a maximum simulation time of $t = 600$ ps was
reached. Because of the long lifetime of the van-der-Waals complexes
formed under initial conditions $E_{\rm coll} = 50$ kcal/mol, $T_{\rm
  rot}=0$ K, these trajectories were only propagated out to 300~ps
because reactions typically occurred within a few ps (see below). \\

\begin{figure}[H] %[htb]
\includegraphics[width=\linewidth]{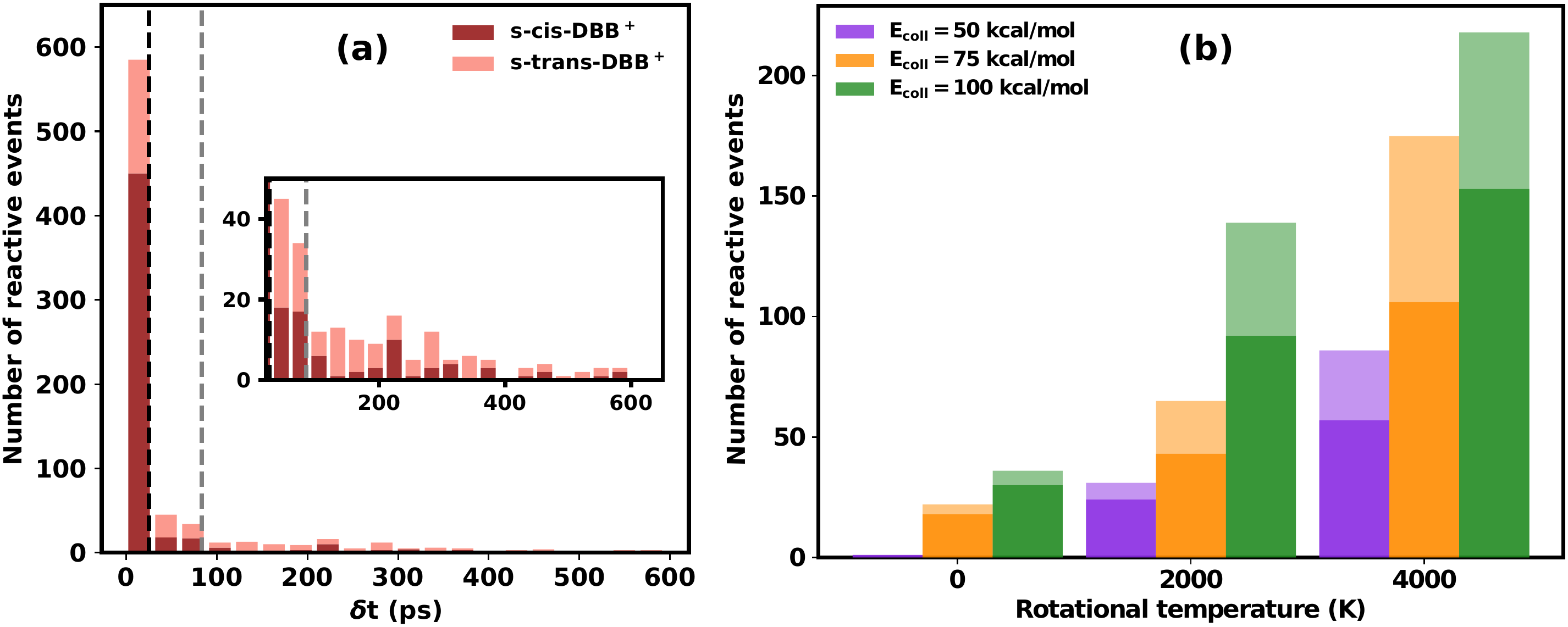}
\caption{(a) Stacked histogram of the
  elapsed time ($\delta t$) for successful reactive
  events. Trajectories starting with s-\textit{cis} and
  s-\textit{trans-}DBB$^+$ are shown in dark and light brown,
  respectively. The mean of the distributions are indicated as dashed
  black and grey vertical lines for s-\textit{cis-}DBB$^+$ and
  s-\textit{trans-}DBB$^+$, respectively. The inset shows a
  magnification of the tail of the distribution. (b) Stacked
  histogram of the variation of the number of reactive events at
  collision energies 50, 75 and 100~kcal/mol with vibrational
  temperature 100~K and impact parameter $b=$0~$\rm \AA$~as a function
  of the rotational temperature of the reactant molecules. Reactive
  events from trajectories that started with s-\textit{cis} and
  s-\textit{trans}-DBB$^+$ are represented in solid and transparent
  colors, respectively.}
\label{fig:figure8}
\end{figure}  	

\noindent   	
As the MDP suggested and Figure \ref{fig:figure8}(b) confirms,
rotational energy promotes the reaction even if the rotational degrees
of freedom are less active for the ionic MDP than for the neutral
reaction\cite{rivero19b}. In fact, there are almost five times more
reactive events at $E_{\rm coll}=75$~kcal/mol, $T_{\rm rot}=4000$~K
than at $E_{\rm coll}=100$~kcal/mol, $T_{\rm rot}=0$~K even though
these scenarios exhibit similar total kinetic energies. \\

\noindent
To assess whether the reaction is direct or complex-mediated, the time
$t_1$ of surface crossing between the reactant and intermediate force
fields is shown in Figure S6 of the SM for all reactive
trajectories. It was found that $t_1< 2$~ps for the majority of
trajectories and that $t_1 < 7$~ps for all reactive trajectories
indicating that they are direct events. Further, the times of surface
crossing for $b \in [0,6]$ are summarized in Figures S7
and S8 of the SM for s-\textit{cis}- and s-\textit{trans}-DBB$^{+}$,
respectively. The majority of these reactive trajectories exhibited a
reaction time between 0.5 to 1~ps. Reaction times longer than 5~ps
have not been observed for either conformer. \\

\noindent
To determine the synchronicity of the reactions, the time $\delta t$
elapsed between formation of the first and second carbon-carbon bond
was calculated for all reactive events as the time difference between
the crossing from the reactant force field to the intermediate force
field ($t_1$) and the crossing from the intermediate force field to
the product force field ($t_2$) (Figure \ref{fig:figure8}(a)),
i.e. $\delta t = t_2 - t_1$. The times $t_1$ and $t_2$ are only
approximate time stamps for the formation of the first and second
bonds since the system crosses surfaces at C-C distances longer than
1.6~\AA~ that is the usual threshold for formation of these type of
bonds. However, if the system remains on the intermediate force field
(for $t_1$) and the product force field (for $t_2$) the bond is formed
because otherwise the system would cross back to the reactant-state
PES. Out of 773 reactive events, only 59 were found to be synchronous
with $\delta t <30$~fs \cite{souza16a, diau99a} (see
Table~\ref{table-cat}). All synchronous processes start from
s-\textit{cis}-DBB$^+$. The intermediate species has lifetimes on the
order of picoseconds. It is important to remember that our model is
expected to somewhat overestimate the lifetime of the intermediate due
to the high activation barriers along the \textit{trans} path (Figure
\ref{fig:figure2}(c)). \\

\noindent	
At the end of the maximum simulation time interval, there were still
$\sim 30000$ van der Waals complexes left from the original sample of
trajectories (see Table~\ref{table-cat}) that could eventually form
products on longer time scales. However, no product formation with
$t_1> 7$~ps was recorded even though some of the van der Waals
complexes live for 600~ps, so this is expected to be an unlikely and
slow process. Nevertheless, the effect of these complexes on the total
rate could be included once a direct comparison with experiment is
possible.\cite{Koner2014,MM.mgo:2020} The 245 trajectories that are
trapped in the intermediate region could eventually evolve to products
or dissociate. However, as can be seen in Figure \ref{fig:figure8}(a), formation of the second bond on the $>100$ ps time scale after
the first bond was formed only occurs in $< 1$ \% of the
cases. Therefore, the contribution of such trajectories to the final
rate is expected to be small.\\

\noindent
After this qualitative overview of the reactive dynamics, the rate of
reaction was estimated from a second set of trajectories by scanning
the impact parameter $b$ over a finite range. For these studies,
$7\cdot10^5$ and $6\cdot10^5$ initial structures for the
s-\textit{cis}-DBB$^{+}$ and s-\textit{trans}-DBB$^{+}$ conformers
were generated, respectively. The impact factor ($b$) was chosen
between 0 and 6 \AA\/ and was uniformly sampled in six non-overlapping
intervals, with increments of 1 \AA\/. For every interval, $10^5$
reactive MD simulations were run. Further, $10^5$ trajectories were
simulated with $b = 0$ \AA\/ for both conformers to connect with the
first set of simulations as described above. The collision energy was
set to 100 kcal/mol, and the trajectories were simulated for 50 ps
with $\Delta t = 0.1$ fs and at 300 K. The two sets of simulations
were found to be consistent with one another as for $b=0$ the fraction
of reactive trajectories from the first set is $4.3\times 10^{-4}$,
compared with $4.6\times 10^{-4}$ from the second set for the same
simulation conditions.\\

\begin{table}[H]
\centering
\caption{Number of reactive trajectories at specific impact parameters $b$ 
  for s-\textit{cis}-DBB$^{+}$ and s-\textit{trans}-DBB$^{+}$.  }\
\begin{tabular}{c c c c c c c c}
\toprule
$b$ / \AA &0&0-1&1-2&2-3&3-4&4-5&5-6 \\
\midrule
s-\textit{cis}-DBB$^{+}$ & 133& 140& 76& 22& 10& 2& 0 \\
s-\textit{trans}-DBB$^{+}$ & 64& 72& 46& 33& 4& 0& -\\
\bottomrule
\end{tabular}
\label{opacity_table}
\end{table}

\begin{figure}[H]
\centering \captionsetup{}
\includegraphics[scale=0.5]{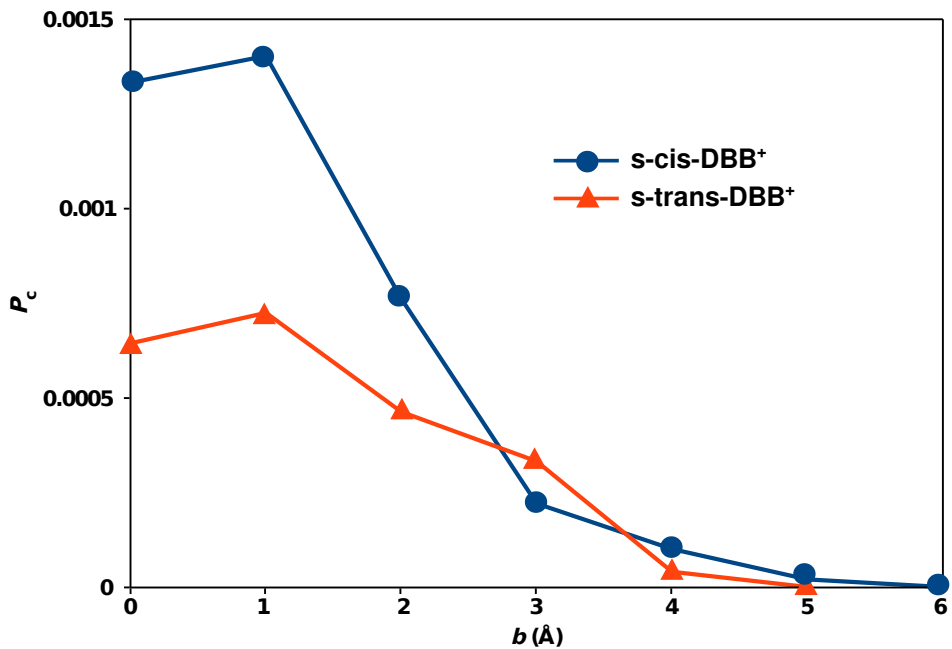}
\caption{Opacity function for the reaction of s-\textit{cis}-DBB$^{+}$
  (blue) and s-\textit{trans}-DBB$^{+}$ (red) with MA as a function of
  impact parameter $b$. }
\label{fig:figure9}
\end{figure}

\noindent
The opacity functions for s-\textit{cis}-DBB$^{+}$ and
s-\textit{trans}-DBB$^{+}$ (see Table~\ref{opacity_table}) are
presented in Figure \ref{fig:figure9}. The s-\textit{cis}-conformer
was found to have a higher reaction probability for all impact
factors, except around $b \sim 3$~\AA\/. For head-on collisions
($b=0$), the number of reactive trajectories was 133 for
s-\textit{cis}- and 64 for s-\textit{trans}-DBB$^{+}$, respectively,
which is consistent with the extended set of dynamics computed for
$b=0$ discussed above. For $b \in [0,1]$~\AA\, the number of reactive
trajectories increases slightly to 140 and 72 for s-\textit{cis}- and
s-\textit{trans}-DBB$^{+}$, respectively. However, this increase may
not be statistically significant considering the still relatively
small numbers of reactive events sampled. For larger impact
parameters, the opacity function decays monotonically to reach zero
around $b \in [5,6]$ \AA\/ for s-\textit{cis}- and around $b \in
[4,5]$ \AA\/ for s-\textit{trans}-DBB$^{+}$. Therefore, no simulations
were carried out for $b \in [5,6]$ \AA\/ for
s-\textit{trans}-DBB$^{+}$. \\

\noindent
The reaction rates for the two conformers were calculated from the
opacity functions. For a uniform sampling, all trajectories were
grouped in non-overlapping intervals of $b$ with a weight\\
\begin{equation}
w = \dfrac{2b}{b_{\rm max}}
\end{equation}
where $b_{\rm max}$ is the maximum value of $b$ for which a reactive
complex is formed, i.e. $b_{\rm max} = 5$ \AA\/ for
s-\textit{trans}-DBB$^{+}$ and $b_{\rm max} = 6$ \AA\/ for
s-\textit{cis}-DBB$^{+}$.

\noindent
The reaction probability in each interval was calculated as
\begin{equation}
P_{\rm c} = \dfrac{N^{'}_{\rm r}}{N_{\rm tot}}.
\end{equation}
Here, $N^{'}_{\rm r}$ is the effective number of reactive
trajectories,
\begin{equation}
N^{'}_{\rm r} = \sum_{i=1}^{N_{\rm r}} w_{i},
\end{equation}
where $N_{\rm r}$ and $N_{\rm tot}$ is the number of reactive and
total trajectories within the specific interval, respectively.

The rate coefficient was determined according to
\begin{equation}
k(T)=\sqrt{\dfrac{8k_{b}T}{\pi \mu}} \time 2 \pi b^{2}_{max}
P_{c}.
\end{equation}
yielding $k = 5.116 \times 10^{-14}$ s$^{-1}$ for
s-\textit{cis}-DBB$^{+}$ and $k = 3.796 \times 10^{-14}$ s$^{-1}$ for
s-\textit{trans}-DBB$^{+}$ at an internal, i.e.,
rotational-vibrational, temperature of 300~K and a collision energy of
100~kcal/mol.\\

\noindent
Comparing with the neutral variant of the reaction, it can be
concluded that the ionic system is considerably more reactive. This is
exemplified by the ratio between the reactive number of trajectories
and the total number trajectories which was found to be $6.79 \times
10^{-5}$ in the neutral system \cite{rivero19b} and $4.29 \times
10^{-4}$ for the ionic reaction studied here, i.e. a difference of
about one order of magnitude. For the neutral reaction comparable
rates as for the ionic system were only obtained at markedly higher
rotational temperatures of $T_\text{rot} = 4000$~K \cite{rivero19b},
while at $T_\text{rot}=300$~K, only a negligibly small number of
reactive events was observed.\\

\section{Conclusions}
The cationic Diels-Alder reaction of maleic anhydride with
2,3-dibromobutadiene ions has been studied using reactive molecular
dynamics. Trajectories were initiated in configurations with the two
reactant molecules approaching each other mimicking a collision
experiment. A competition of concerted and stepwise reaction pathways
was found and both, the s-\textit{cis} and s-\textit{trans} conformers
of the diene proved to be reactive. These findings are in contrast
with the usual paradigm assumed for neutral Diels-Alder reactions as
concerted processes in which only the s-\textit{cis} conformer of the
diene can react. The analysis of the minimum dynamic path of the
reaction indicates that both, rotations and vibrations are important
to drive the system towards the transition state, whereas for the
neutral reaction\cite{rivero19b} only rotations were found to play an
important role in promoting the reaction. This may be rationalized by
the fact that the transition state of the cationic concerted reaction
pathway is asymmetric, whereas the one of the neutral variant is
symmetric. Because for the ionic reaction the reactant molecules are symmetric, deformations of the molecular structures along the
reaction path are more pronounced and vibrations are more highly excited  along
the MDP. This analysis of the minimum dynamic
path clarified the role of rotational and vibrational degrees of
freedom which provides valuable information for the design of future
experiments.\\

\noindent
Another difference between the two types of reactions is that the
cationic system is predicted to form van-der-Waals complexes even at
the high collision energies of 50, 75, and 100 kcal/mol considered
here. However, at the energies at which reactive events were recorded
in the present study, the reactions were found to be direct and mostly
asynchronous although some cases of synchronous trajectories were also
identified. This underlines that computationally efficient energy functions are mandatory that allow running a statistically significant number of reactive trajectories such as to cover a broad range of possible scenarios. The ionic system was found to be more reactive than its
neutral counterpart in line with the difference in activation energies
of the two systems.\\

\noindent
Although the accuracy of the present MS-ARMD PES is moderate compared
with what is possible by using neural network\cite{MM.physnet:2019} or
kernel-based PESs,\cite{MM.rkhs:2017,tkatchenko:2018,MM.rkhsf:2020} it
needs to be stressed that running a statistically meaningful number of
trajectories (here $10^6$ to $10^7$) for a system of the present size
is currently only viable with a force field-inspired technique such as
MS-ARMD. Whenever rates and quantities derived from simulations using
two such different approaches have been made, they agree closely,
though.\cite{rivero19b,MM.mgo:2020,MM.h2co:2020} Hence, although
quantitative aspects of the MS-ARMD PES can be further improved, the
qualitative conclusions about the reaction dynamics of the present
reaction are deemed to be correct.\\

\noindent
The present study highlights salient dynamic differences between
neutral and ionic Diels-Alder reactions and represents a stepping
stone towards a rigorous investigation of their dynamics in
conformationally controlled gas-phase experiments \cite{willitsch17a,
  chang13a}.\\

\section*{Acknowledgment}
Support by the Swiss National Science Foundation through grants
BSCGI0\_157874 (to SW), 200021\_117810, 200020\_188724, and the NCCR MUST (to MM), and the
University of Basel is acknowledged. \\

\bibliography{references} 

\end{document}

\doublespace

%%%%%%%%%% PRELIMINARY MATERIAL %%%%%%%%%%
\maketitle
\thispagestyle{empty}
%%%%%%%%%% MAIN TEXT STARTS HERE %%%%%%%%%%
\begin{abstract}
	
\end{abstract}
\maketitle

\section{Supplementary figures}

\begin{figure}[H]
\includegraphics[width=0.65\textwidth]{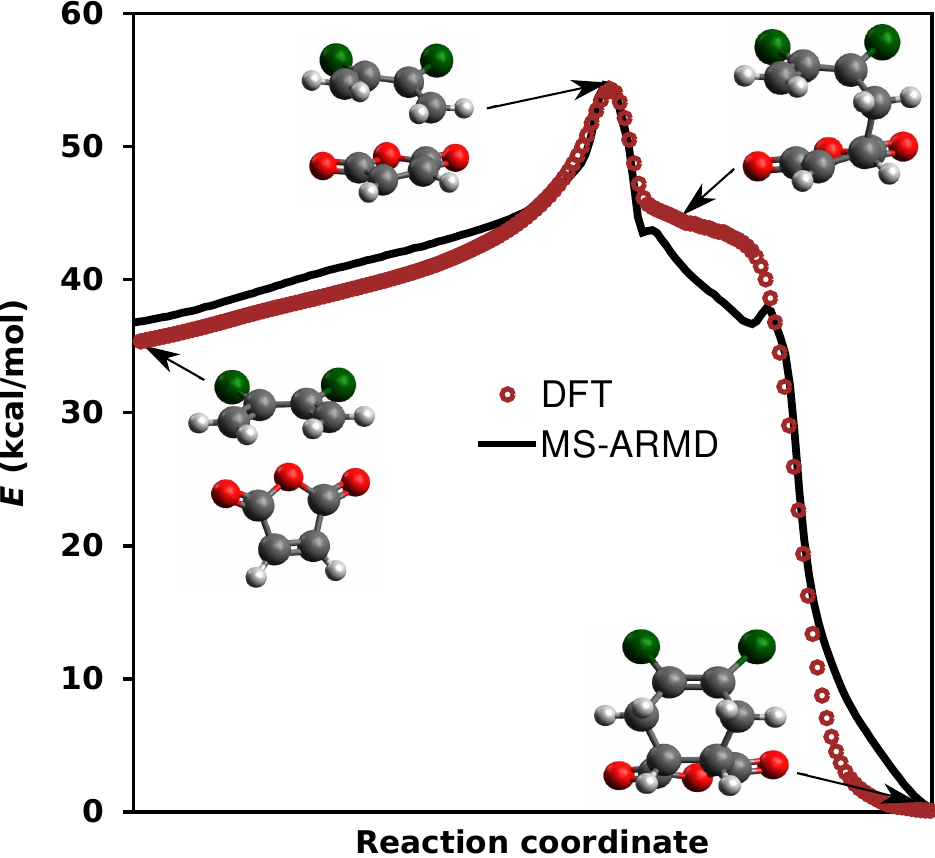}
\caption[Endo IRC at the M06-2X/6-31G* level of
  theory]{\textit{Endo}-IRC at the M06-2X/6-31G* level of theory with
  some relevant structures along the path.}
\label{figureS1}
\end{figure}

\begin{figure}[H]
\includegraphics[width=.9\textwidth]{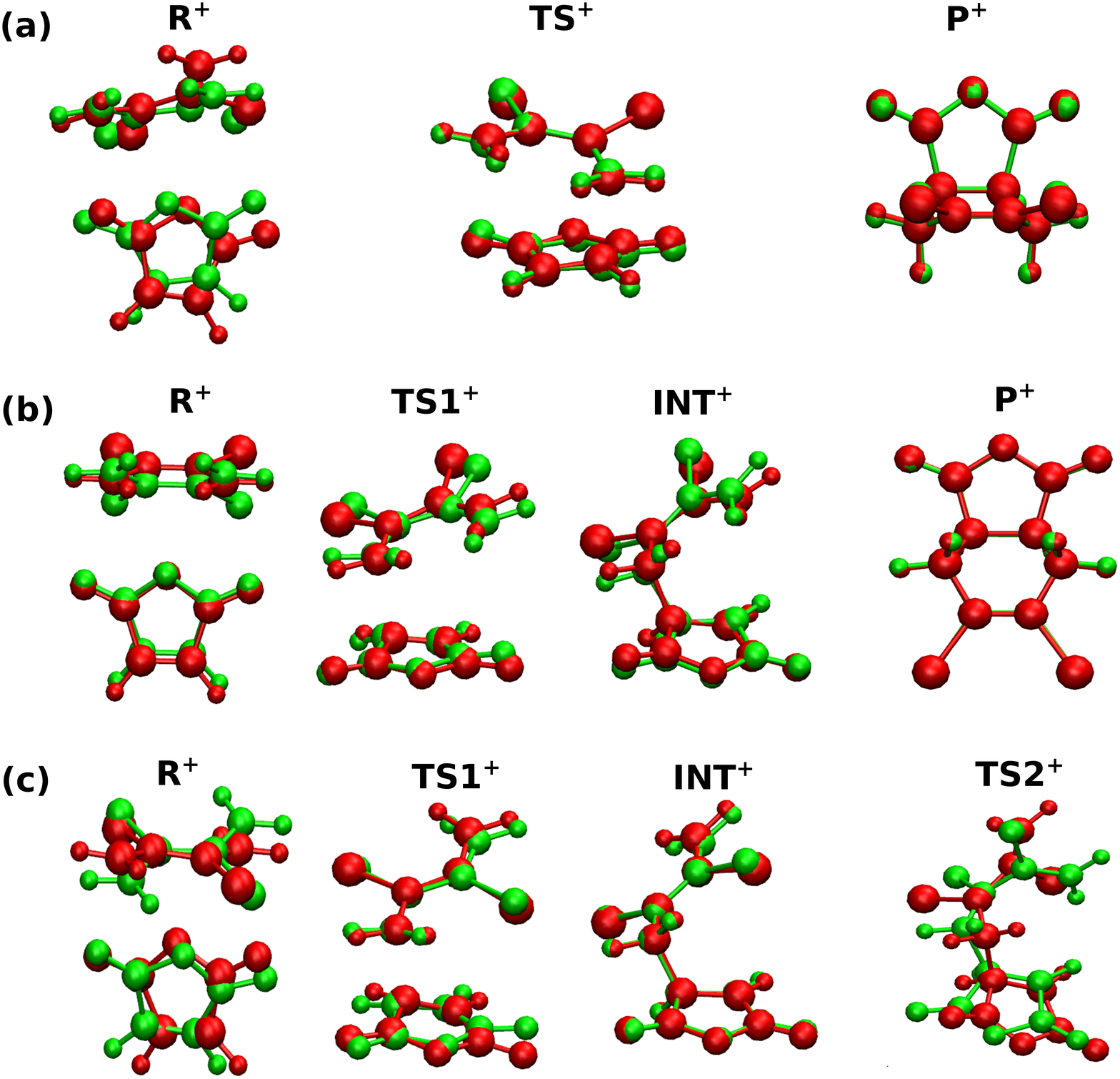}
\caption{Comparison of optimized stationary-point structures along the
  (a) \textit{endo}, (b) \textit{exo} and (c) \textit{trans} paths at
  the DFT (red) and MS-ARMD (green) levels of theory.}
\label{figureS2}
\end{figure}

\begin{table}[H]
\caption{RMSD between the DFT and MS-ARMD optimized stationary-point structures along the \textit{endo},  \textit{exo} and \textit{trans} paths. } 
\begin{tabular}{c|c}
Molecules & RMSD (\AA\/) \\ 
\hline
MA  & 0.003    \\
cis-DBB$^+$ & 0.036   \\
trans-DBB$^+$ & 0.083 \\
\hline
Complexes & RMSD (\AA\/) \\ 
\hline
TS$^+$ (\textit{endo}) & 0.325  \\
P$^+$ (\textit{endo}) & 0.073 \\
TS1$^+$ (\textit{exo}) & 0.236 \\
INT$^+$ (\textit{exo}) & 0.186 \\
P$^+$ (\textit{exo}) & 0.074 \\
TS1$^+$ (\textit{trans}) & 0.270 \\
INT$^+$ (\textit{trans}) & 0.101 \\
TS2$^+$ (\textit{trans}) & 0.328 \\
\end{tabular} 
\label{tabRMSD}
\end{table}

\begin{figure}[H]
\includegraphics[width=\textwidth]{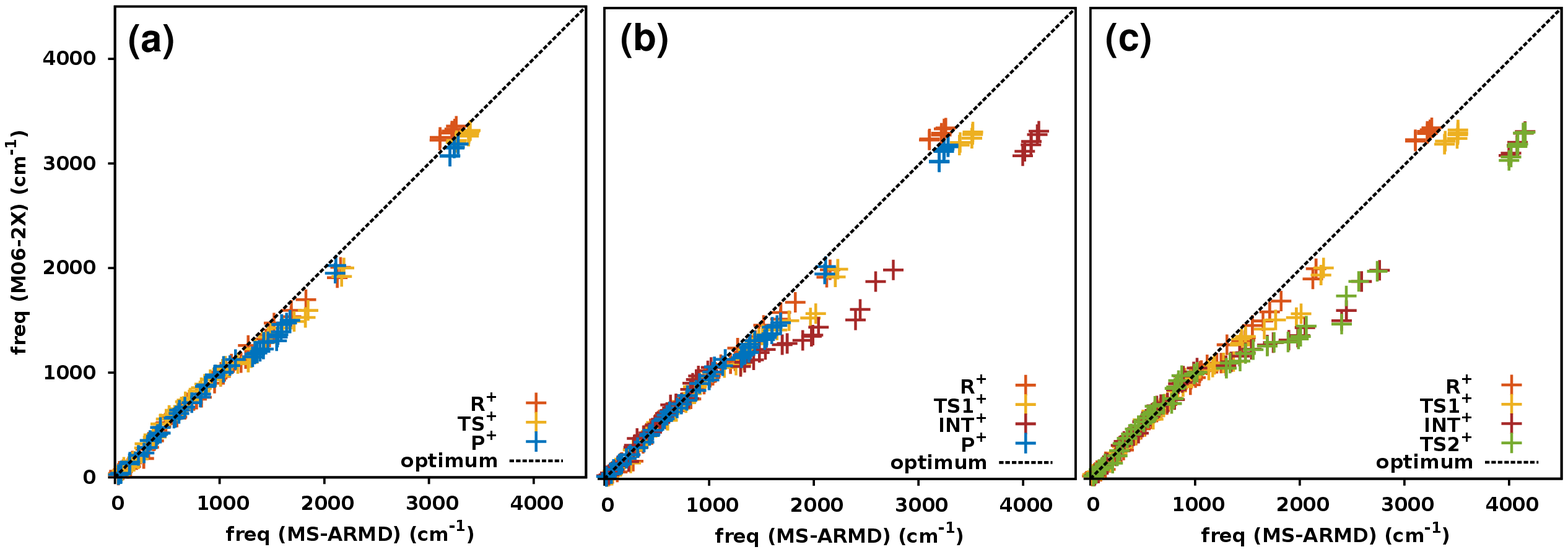}
\caption{Comparison of harmonic frequencies at the DFT level of theory
  and from the MS-ARMD PES for the minimized structures along the (a)
  \textit{endo}, (b) \textit{exo} and (c) \textit{trans} paths in
  Figure \ref{figureS2}}
\label{figureS3}
\end{figure}

\begin{figure}[H]
\includegraphics[width=\textwidth]{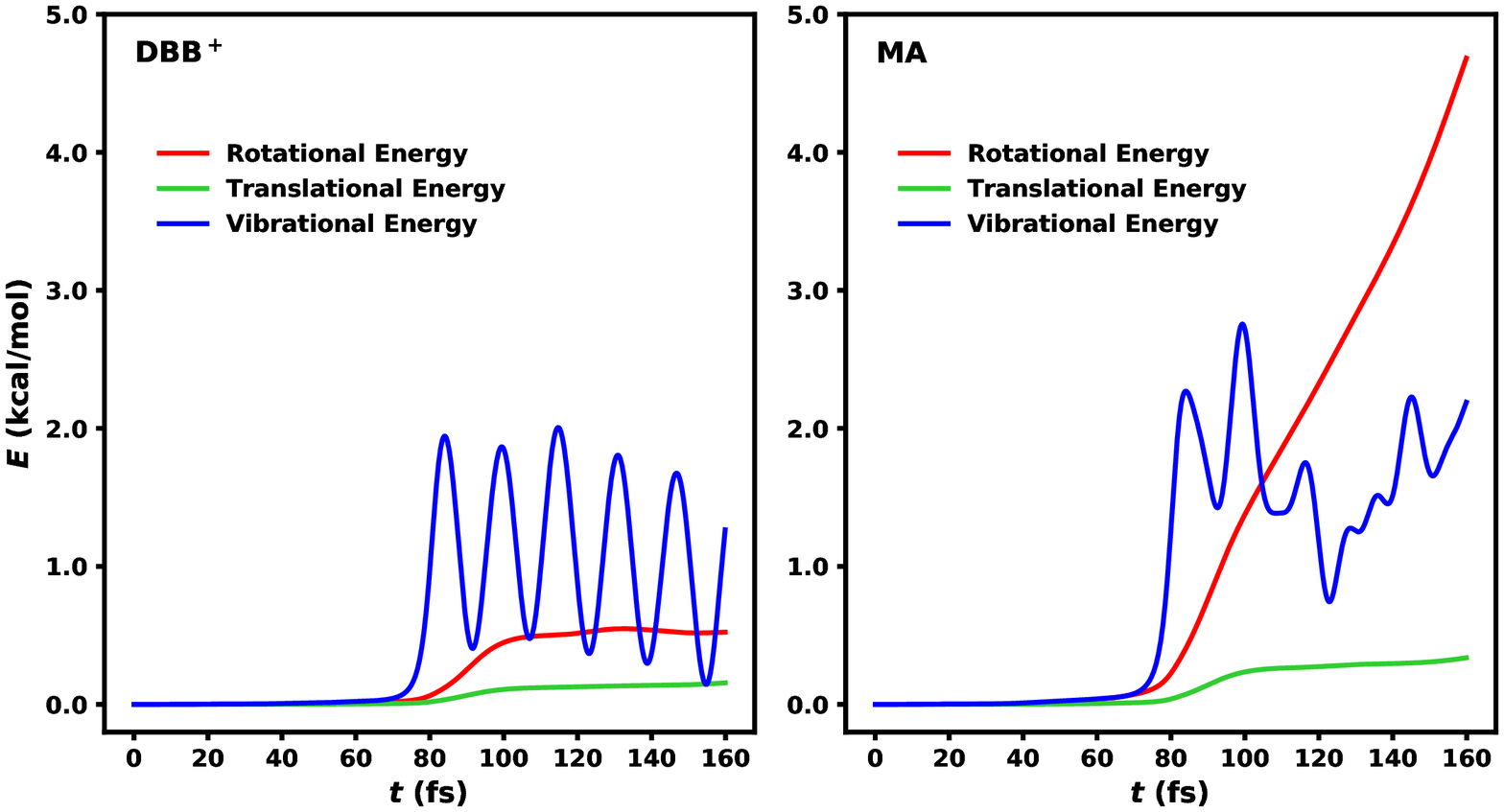}
\caption{Decomposition of the total kinetic energy ($E$) along the
  minimum dynamic path into rotational, translational and vibrational
  energy. The trajectories start at TS1-exo$^+$ and ends at the
  reactants. }
\label{figureS4}
\end{figure}

\begin{figure}[H]
\includegraphics[width=\textwidth]{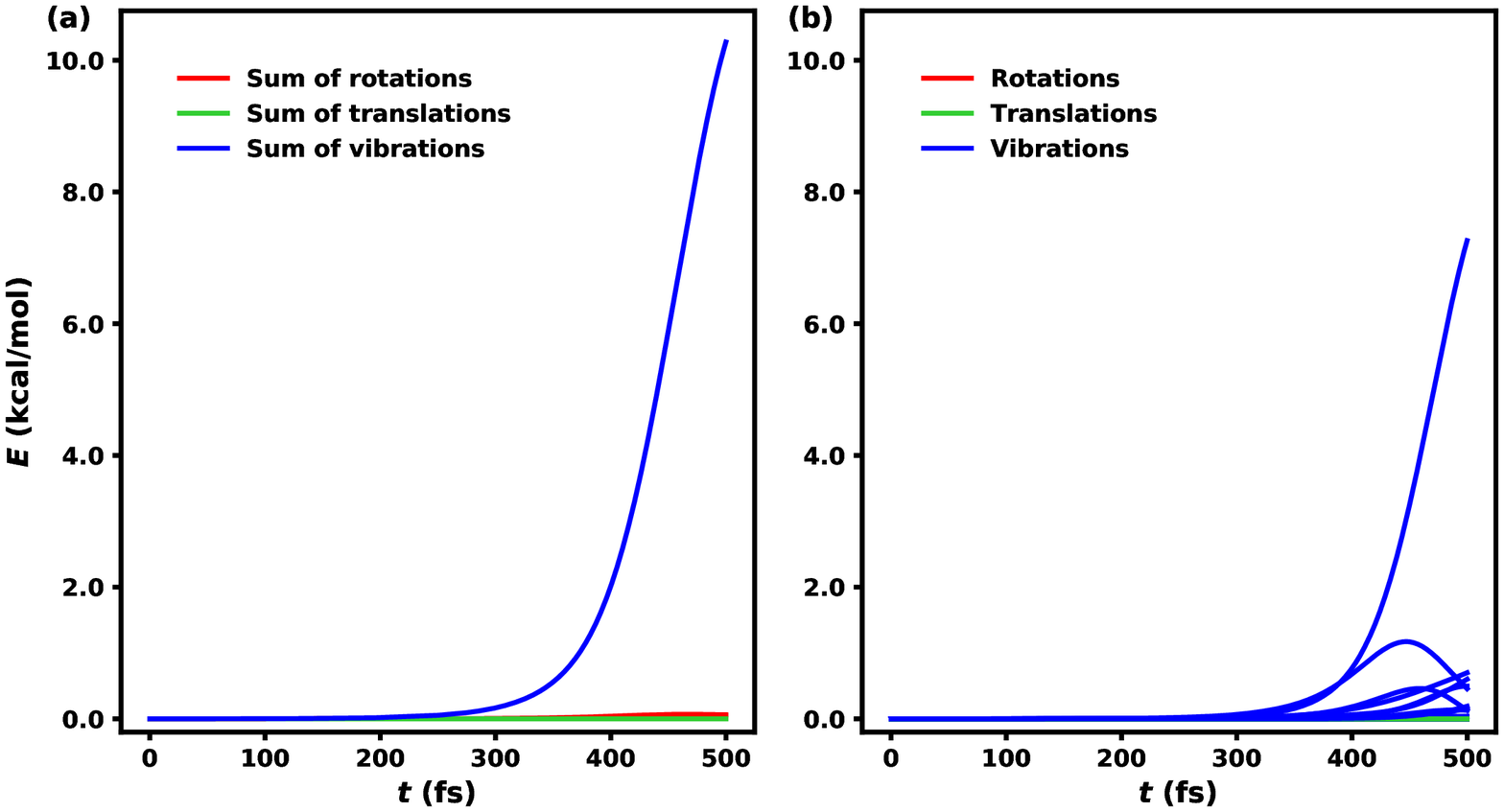}
\caption{Projection of the total kinetic energy ($E$) onto the degrees
  of freedom of INT-tr$^+$ along the minimum dynamic path for the
  \textit{cis/trans} isomerization of the intermediate (a) summed into
  rotations, translations and vibrations (b) as individual traces. The
  trajectories start at TS2-trans$^+$ and ends at INT-tr$^+$. The most
  active vibration is the \textit{cis/trans} isomerization mode.}
\label{figureS5}
\end{figure}

\begin{figure}[H]
\includegraphics[width=0.47\textwidth]{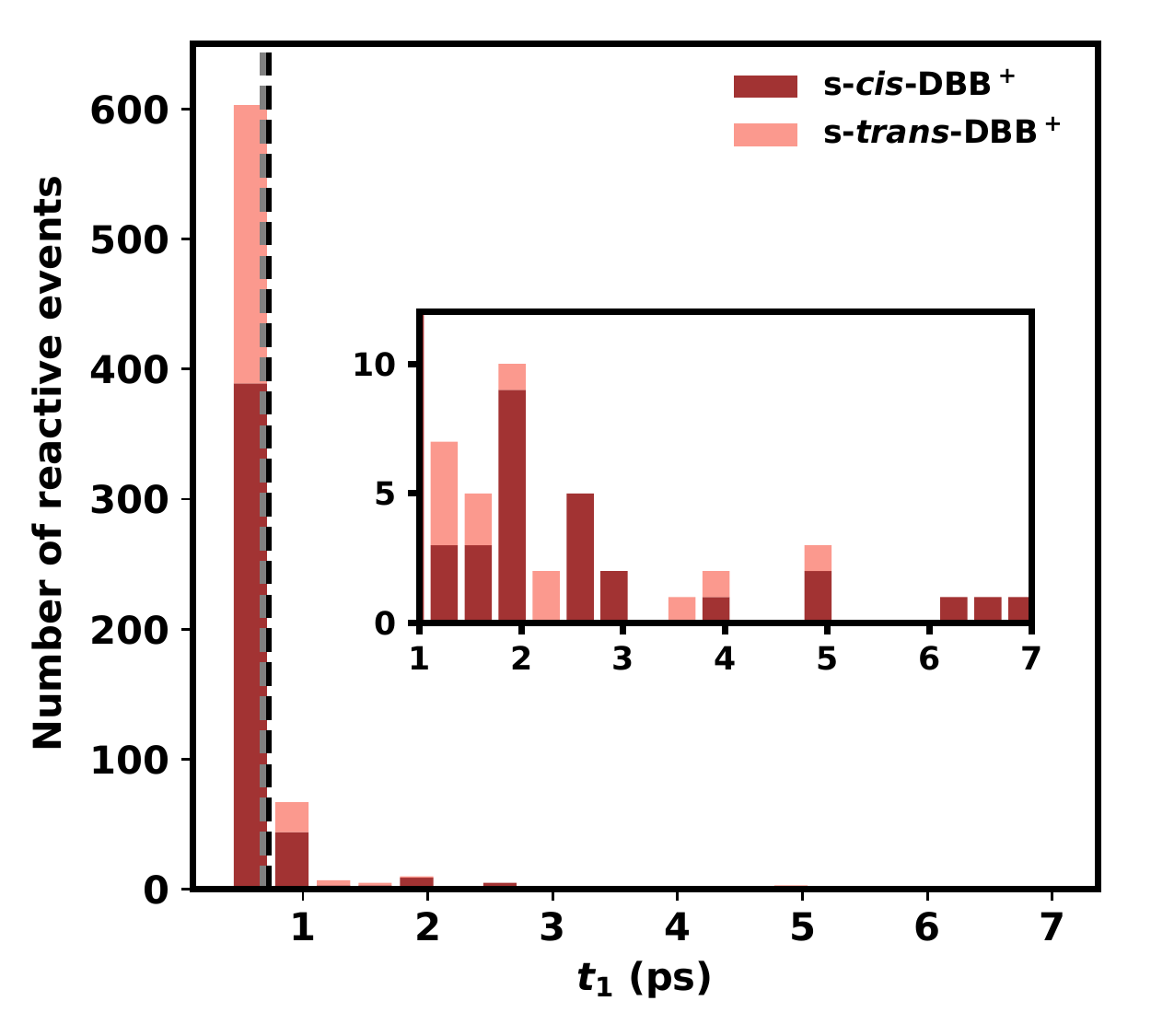}
\caption[Synchronicity of the reactive events and time of formation of
  the first bond]{Stacked histogram of the time of formation of the first bond ($t_1$) for all
  reactive events. Trajectories starting with s-\textit{cis} and
  s-\textit{trans}-DBB are displayed in dark and light brown,
  respectively. The mean of the distributions are indicated as
  vertical lines in black for s-\textit{cis}-DBB$^+$ and in gray for
  s-\textit{trans}-DBB$^+$. The inset shows a magnification of the
  tail of the distribution.}
 \label{figureS6}
 \end{figure}  
 		
 \begin{figure}[H]
\centering
\includegraphics[scale=0.48]{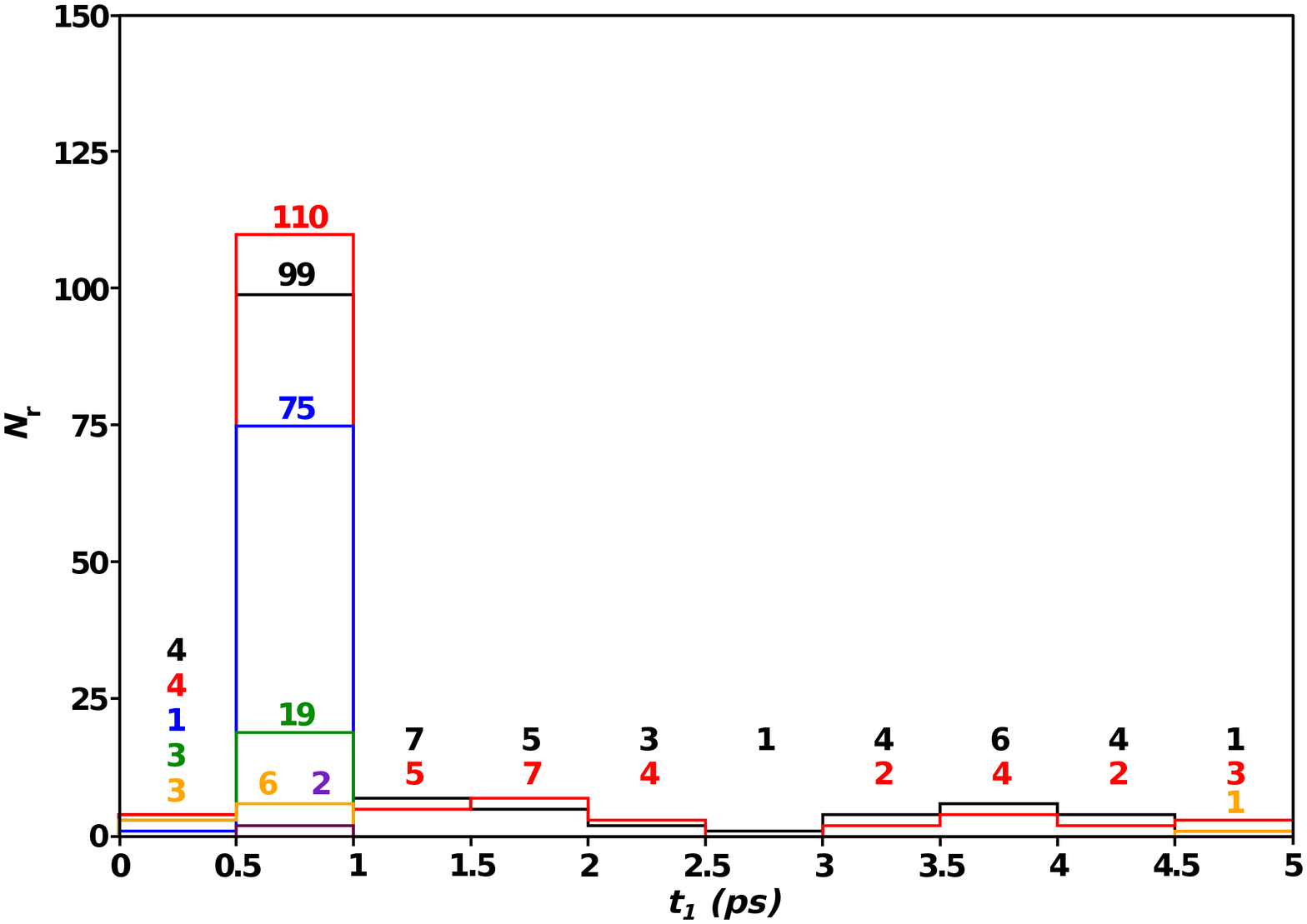}
\caption{Stacked histogram of the formation time for the first bond
  ($t_1$) for s-\textit{cis}-DBB$^{+}$ with binning of 0.5 ps. Color
  code: $b = 0$ (black), $b \in [0,1]$ (red), $b \in [1,2]$ (blue), $b
  \in [2,3]$ (green), $b \in [3,4]$ (orange), $b \in [4,5]$
  (violet). The values on top of the bins represent the number of
  reactive trajectories within the relevant time intervals.}
\label{figureS7}
\end{figure}

\begin{figure}[H]
\centering
\includegraphics[scale=0.48]{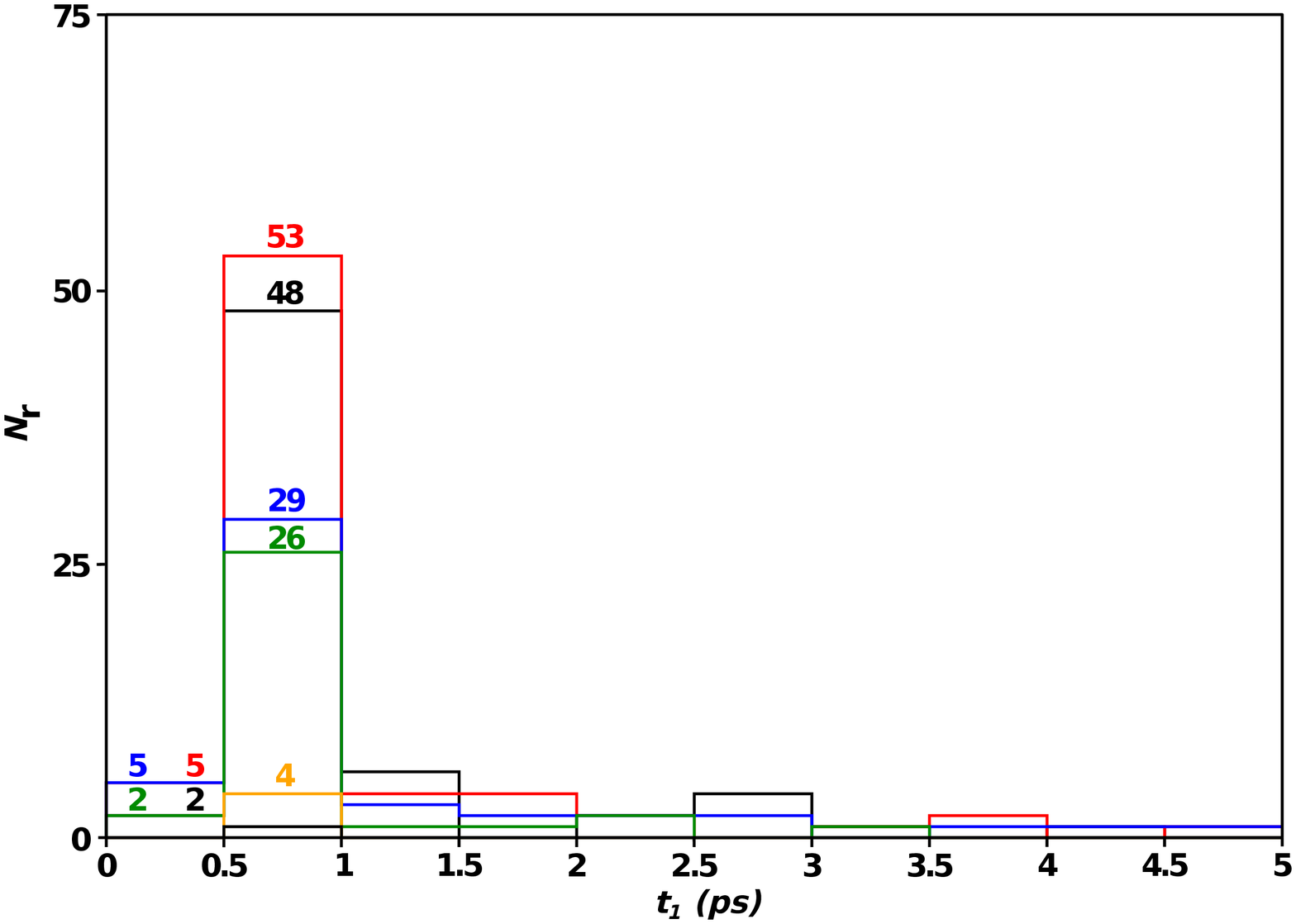}
\caption{Stacked histogram of the formation time for the first bond
  ($t_1$) for s-\textit{trans}-DBB$^{+}$ with binning of 0.5 ps. Color
  code: $b = 0$ (black), $b \in [0,1]$ (red), $b \in [1,2]$ (blue), $b
  \in [2,3]$ (green), $b \in [3,4]$ (orange). The values on top of the
  bins represent the number of reactive trajectories within the
  relevant time intervals.}
\label{figureS8}
\end{figure}

\section*{Parametrization of the MS-ARMD PES for the Diels-Alder reaction DBB$^+$~+~MA}

  \begin{table}[H]
  \footnotesize
    \caption{Harmonic bond parameters of the MS-ARMD force
      fields. $k/2$ is given in units of kcal/mol/\AA$^2$ and $r_e$ is
      in \AA.}  \resizebox{\textwidth}{!}{
  \begin{tabular}{c|c|c|c|c|c|c|c}
  \multicolumn{2}{c|}{ Atoms }& \multicolumn{2}{c|}{Reactant FF}& \multicolumn{2}{c|}{ Intermediate FF} &\multicolumn{2}{c}{Product FF} \\ \hline
 1 \# & 2 \# & $k$/2 & $r_e$ & $k$/2  & $r_e$ & $k$/2  & $r_e$ \\ \hline
2  &  4   &    1093.65  & 1.19577        &        1398.03  & 1.19312    &       1029.89  & 1.19351 \\
3  & 5    &    1093.65  & 1.19577        &        1398.03  & 1.19312    &       1029.89  & 1.19351 \\
6  & 8    &    411.539  & 1.08513        &        650.507  & 1.08709    &       411.200  & 1.09308 \\
7  & 9    &    411.539  & 1.08513        &        650.507  & 1.08709    &       411.200  & 1.09308 \\
11 & 15   &    391.619  & 1.08580        &        650.507  & 1.08709    &       411.200  & 1.09308 \\
12 & 17   &    391.619  & 1.08580        &        650.507  & 1.08709    &       411.200  & 1.09308 \\
11 & 13   &    391.619  & 1.08580        &        650.507  & 1.08709    &       411.200  & 1.09308 \\
10 & 12   &    391.619  & 1.08580        &        650.507  & 1.08709    &       411.200  & 1.09308 \\

  \end{tabular} }
  \label{tabS1}
  \end{table}

\begin{table}[H]
\caption{Morse bond parameters for the MS-ARMD PESs. ``X'' indicates
  that this parameter is not needed. $D_e$ is in kcal/mol, $r_e$ in
  \AA~and $\beta$ in \AA$^{-1}$. } \resizebox{\textwidth}{!}{
  \begin{tabular}{c|c|c|c|c|c|c|c|c|c|c}
  \multicolumn{2}{c|}{ Atoms }& \multicolumn{3}{c|}{Reactant FF}& \multicolumn{3}{c|}{ Intermediate FF} &\multicolumn{3}{c}{ Product FF} \\ \hline
 1 \# & 2 \# & $D_e$ & $r_e$ & $\beta$  & $D_e$  & $r_e$ & $\beta$ & $D_e$  & $r_e$ & $\beta$\\ \hline
14  &  18   &    187.087 &  1.83777 &  1.10111  &   85.9534 &  1.81583 &  1.20270 &    91.2590 &  1.80871 &  1.89304 \\ 
16  &  19   &    187.087 &  1.83777 &  1.10111  &   85.9534 &  1.81583 &  1.20270 &    91.2590 &  1.80871 &  1.89304 \\ 
1   & 2     &    85.1832 &  1.39166 &  1.99423  &   101.912 &  1.34206 &  1.87376 &    474.144 &  1.37373 &  1.00021 \\ 
1   & 3     &    85.1832 &  1.39166 &  1.99423  &   101.912 &  1.34206 &  1.87376 &    474.144 &  1.37373 &  1.00021 \\ 
2   & 6     &    164.953 &  1.50368 &  1.40025  &   63.8103 &  1.55906 &  1.18897 &    189.327 &  1.52650 &  1.38405 \\ 
3   & 7     &    164.953 &  1.50368 &  1.40025  &   63.8103 &  1.55906 &  1.18897 &    189.327 &  1.52650 &  1.38405 \\ 
14  &  16   &    564.329 &  1.40983 &  0.896631 &   327.948 &  1.41861 &  1.44211 &    310.137 &  1.42532 &  1.12287 \\ 
11  &  14   &    189.781 &  1.36281 &  1.63991  &   76.1220 &  1.32884 &  1.01309 &    236.306 &  1.49451 &  1.24374 \\ 
12  & 16    &   189.781  & 1.36281  & 1.63991   &   76.1220 &  1.32884 &  1.01309 &    236.306 &  1.49451 &  1.24374 \\ 
6   & 7     &    194.950 &  1.33603 &  1.99291  &   154.586 &  1.50359 &  1.94850 &    58.3539 &  1.47745 &  1.99994 \\ 
7   & 12    &       X    &     X    &     X     &   116.127 &  1.51263 &  1.01790 &    58.3539 &  1.47745 &  1.99994 \\ 
6   & 11    &       X    &     X    &     X     &     X     &     X    &     X    &    58.3539 &  1.47745 &  1.99994 \\

  \end{tabular} }
  \label{tabS2}
  \end{table}

\begin{table}[H]
  	\scriptsize
    \caption{Angle parameters for the MS-ARMD PESs. ``X'' indicates
      that this parameter is not needed. $k$/2 is in
      kcal/mol/radian$^2$, $\theta_e$ in degree}
    \resizebox{\textwidth}{!}{
  \begin{tabular}{c|c|c|c|c|c|c|c|c}\multicolumn{3}{c|}{ Atoms }& \multicolumn{2}{c|}{Reactant FF}& \multicolumn{2}{c|}{ Intermediate FF}& \multicolumn{2}{c}{ Product FF}\\ \hline 
  1 \# & 2 \# & 3 \# & $k$/2 & $\theta_e$ & $k$/2 & $\theta_e$ & $k$/2 & $\theta_e$\\\hline
      2  &  1   & 3        &   64.5319  & 105.978   &  135.149  & 97.7709    &  109.386  & 111.098  \\ 
      1  &  2   & 4        &   99.6847  & 128.737   &  165.070  & 130.560    &  114.817  & 135.968  \\ 
      1  &  2   & 6        &   135.436  & 109.951   &  277.466  & 107.951    &  134.612  & 120.559  \\ 
      4  &  2   & 6        &   77.1982  & 137.451   &  78.2342  & 143.084    &  71.8608  & 148.445  \\ 
      1  &  3   & 5        &   99.6847  & 128.737   &  165.070  & 130.560    &  114.817  & 135.968  \\ 
      1  &  3   & 7        &   135.436  & 109.951   &  277.466  & 107.951    &  134.612  & 120.559  \\ 
      5  &  3   & 7        &   77.1982  & 137.451   &  78.2342  & 143.084    &  71.8608  & 148.445  \\ 
      2  &  6   & 7        &   95.5494  & 108.805   &  134.514  & 110.650    &  89.4152  & 108.264  \\ 
      2  &  6   & 8        &   38.5239  & 124.711   &  67.2617  & 121.288    &  61.0991  & 109.775  \\ 
      2  &  6   & 11       &     X      &   X       &      X    &     X      &  89.4152  & 108.264  \\ 
      7  &  6   & 8        &   20.1303  & 129.751   &  64.9256  & 125.690    &  54.6427  & 108.583  \\ 
      7  &  6   & 11       &     X      &   X       &      X    &     X      &  82.9790  & 98.3030  \\ 
      8  &  6   & 11       &     X      &   X       &      X    &     X      &  54.6427  & 108.583  \\ 
      3  &  7   & 6        &   95.5494  & 108.805   &  134.514  & 110.650    &  89.4152  & 108.264  \\ 
      3  &  7   & 9        &   38.5239  & 124.711   &  67.2617  & 121.288    &  61.0991  & 109.775  \\ 
      3  &  7   & 12       &     X      &   X       &  134.514  & 110.650    &  89.4152  & 108.264  \\ 
      6  &  7   & 9        &   20.1303  & 129.751   &  64.9256  & 125.690    &  54.6427  & 108.583  \\ 
      6  &  7   & 12       &     X      &   X       &  35.4515  & 133.262    &  82.9790  & 98.3030  \\ 
      9  &  7   & 12       &     X      &   X       &  64.9256  & 125.690    &  54.6427  & 108.583  \\ 
      6  &  11  &  13      &     X      &   X       &     X     &    X       &  54.6427  & 108.583  \\ 
      6  &  11  &  14      &     X      &   X       &     X     &    X       &  65.2956  & 104.164  \\ 
      6  &  11  &  15      &     X      &   X       &     X     &    X       &  54.6427  & 108.583  \\ 
      13 &   11 &   14     &   44.8358  & 132.443   &  123.161  & 121.235    &  58.4988  & 108.370  \\ 
      13 &   11 &   15     &   30.0806  & 138.270   &  106.624  & 119.564    &  44.9477  & 107.173  \\ 
      14 &   11 &   15     &   44.8358  & 132.443   &  123.161  & 121.235    &  58.4988  & 108.370  \\ 
      7  &  12  &  10      &     X      &   X       &  64.9256  & 125.690    &  54.6427  & 108.583  \\ 
      7  &  12  &  16      &     X      &   X       &  46.2212  & 134.655    &  65.2956  & 104.164  \\ 
      7  &  12  &  17      &     X      &   X       &  64.9256  & 125.690    &  54.6427  & 108.583  \\ 
      10 &   12 &   16     &   44.8358  & 132.443   &  123.161  & 121.235    &  58.4988  & 108.370  \\ 
      10 &   12 &   17     &   30.0806  & 138.270   &  106.624  & 119.564    &  44.9477  & 107.173  \\ 
      16 &   12 &   17     &   44.8358  & 132.443   &  123.161  & 121.235    &  58.4988  & 108.370  \\ 
      11 &   14 &   16     &   112.001  & 124.423   &  57.8435  & 132.115    &  92.0564  & 132.027  \\ 
      11 &   14 &   18     &   43.8118  & 136.630   &  46.9468  & 138.569    &  51.8620  & 130.256  \\ 
      16 &   14 &   18     &   50.3729  & 127.032   &  22.9482  & 148.260    &  81.7961  & 128.923  \\ 
      12 &   16 &   14     &   112.001  & 124.423   &  57.8435  & 132.115    &  92.0564  & 132.027  \\ 
      12 &   16 &   19     &   43.8118  & 136.630   &  46.9468  & 138.569    &  51.8620  & 130.256  \\ 
      14 &   16 &   19     &   50.3729  & 127.032   &  22.9482  & 148.260    &  81.7961  & 128.923  \\ 
             
  \end{tabular} }
  \label{tabS3}
  \end{table} 

\begin{longtable}[H] {c|c|c|c|c|c|c|c|c}
\caption{Dihedral parameters of the MS-ARMD PESs.``X'' indicates that
  this parameter is not needed. $k$ is in kcal/mol and $\phi$ in
  degree. } \\

\hline \multicolumn{4}{c|}{ Atoms }& & Reactant FF & Intermediate FF & Product FF& {}\\ \hline 
1 \# & 2 \# & 3 \# & 4 \# & N & $k$ & $k$ & $k$ & $\phi$ \\ \hline     
\endfirsthead
    
\hline \multicolumn{4}{c|}{ Atoms }& & Reactant FF & Intermediate FF & Product FF& {}\\ \hline 
1 \# & 2 \# & 3 \# & 4 \# & N & $k$ & $k$ & $k$ & $\phi$ \\ \hline     
\endhead

\endfoot

\hline
\endlastfoot    
    
         1   & 2   & 6  &  7       & 1  &       X       &         -4.14146   &        -0.523798    &      0.00     \\ 
         1   & 2   & 6  &  7       & 2  &    4.79788    &         8.55575    &        1.72536      &    180.00     \\ 
         1   & 2   & 6  &  7       & 3  &       X       &         -1.81563   &        0.267926     &    0.00       \\ 
         1   & 2   & 6  &  8       & 2  &    0.249864   &         2.35984    &        2.15011      &    180.00     \\ 
         1   & 2   & 6  &  8       & 3  &       X       &         0.221796   &        1.66411      &   0.00        \\ 
         1   & 2   & 6  &  11      & 1  &       X       &              X     &        -0.523798    &      0.00     \\ 
         1   & 2   & 6  &  11      & 2  &       X       &              X     &        1.72536      &    180.00     \\ 
         1   & 2   & 6  &  11      & 3  &       X       &              X     &        0.267926     &    0.00       \\ 
         1   & 3   & 7  &  6       & 1  &       X       &        -4.14146    &        -0.523798    &      0.00     \\ 
         1   & 3   & 7  &  6       & 2  &    4.79788    &         8.55575    &        1.72536      &    180.00     \\ 
         1   & 3   & 7  &  6       & 3  &       X       &        -1.81563    &        0.267926     &    0.00       \\ 
         1   & 3   & 7  &  9       & 2  &     0.249864  &         2.35984    &        2.15011      &    180.00     \\ 
         1   & 3   & 7  &  9       & 3  &       X       &         0.221796   &        1.66411      &   0.00        \\ 
         1   & 3   & 7  &  12      & 1  &       X       &        -4.14146    &        -0.523798    &      0.00     \\ 
         1   & 3   & 7  &  12      & 2  &       X       &         8.55575    &        1.72536      &    180.00     \\ 
         1   & 3   & 7  &  12      & 3  &       X       &        -1.81563    &        0.267926     &    0.00       \\ 
         2   & 1   & 3  &  5       & 1  &    0.701169   &        -5.58519    &        -0.126467    &     0.00      \\ 
         2   & 1   & 3  &  5       & 2  &    6.21522    &        -9.49140    &        -5.49755     &    180.00     \\ 
         2   & 1   & 3  &  5       & 3  &    0.298850   &        12.7562     &        3.34467      &    0.00       \\ 
         2   & 1   & 3  &  7       & 2  &    7.17573    &        -1.12647    &        4.91803      &   180.00      \\ 
         2   & 6   & 7  &  3       & 1  &       X       &      -3.95005    &        -1.29628     &    0.00       \\   
         2   & 6   & 7  &  3       & 2  &    8.19928    &        -1.12703    &             X       &      X        \\
         2   & 6   & 7  &  3       & 3  &       X       &      0.390811    &        0.330764     &     0.00      \\   
         2   & 6   & 7  &  9       & 1  &       X       &        1.45039     &        0.336052     &     0.00      \\ 
         2   & 6   & 7  &  9       & 2  &    6.34212    &       -0.211970    &        -2.95419     &    180.00     \\ 
         2   & 6   & 7  &  12      & 1  &       X       &       -5.40684     &        -3.89326     &    0.00       \\ 
         2   & 6   & 7  &  12      & 2  &       X       &        0.454822    &        -2.88140     &     180.00    \\ 
         2   & 6   & 7  &  12      & 3  &       X       &       -0.0439145   &        -0.444273    &     0.00      \\ 
         2   & 6   & 11 &   13     & 1  &       X       &            X       &        0.336052     &     0.00      \\ 
         2   & 6   & 11 &   13     & 2  &       X       &            X       &        -2.95419     &    180.00     \\ 
         2   & 6   & 11 &   14     & 3  &       X       &            X       &        0.035506     &    0.00   \\     
         2   & 6   & 11 &   15     & 1  &       X       &            X       &        0.336052     &     0.00      \\ 
         2   & 6   & 11 &   15     & 2  &       X       &            X       &        -2.95419     &    180.00     \\ 
         3   & 1   & 2  &  4       & 1  &    0.701169   &        -5.58519    &        -0.126467    &     0.00      \\ 
         3   & 1   & 2  &  4       & 2  &    6.21522    &        -9.49140    &        -5.49755     &    180.00     \\ 
         3   & 1   & 2  &  4       & 3  &    0.298850   &        12.7562     &        3.34467      &    0.00       \\  
         3   & 1   & 2  &  6       & 2  &    7.17573    &        12.7562     &        3.34467      &   180.00      \\  
         3   & 7   & 6  &  8       & 1  &       X       &        1.45039     &        0.336052     &     0.00       \\ 
         3   & 7   & 6  &  8       & 2  &    6.34212    &       -0.211970    &        -2.95419     &    180.00      \\ 
         3   & 7   & 6  &  11      & 1  &       X       &          X         &        -3.89326     &    0.00        \\ 
         3   & 7   & 6  &  11      & 2  &       X       &          X         &        -2.88140     &     180.00     \\ 
         3   & 7   & 6  &  11      & 3  &       X       &          X         &        -0.444273    &     0.00       \\ 
         3   & 7   & 12 &   10     & 1  &       X       &        1.45039     &         0.336052    &      0.00      \\ 
         3   & 7   & 12 &   10     & 2  &       X       &        -0.211970   &        -2.95419     &    180.00      \\ 
         3   & 7   & 12  &  16     & 3  &       X       &        1.51659     &         0.035506    &     0.00   \\     
         3   & 7   & 12  &  17     & 1  &       X       &        1.45039     &         0.336052    &      0.00      \\ 
         3   & 7   & 12  &  17     & 2  &       X       &        -0.211970   &        -2.95419     &    180.00      \\ 
         4   & 2   & 6   & 7       & 1  &       X       &        -9.15296    &        -2.89358     &    0.00        \\ 
         4   & 2   & 6   & 7       & 2  &   -0.999624   &       -1.25200    &           1.75806     &    180.00      \\
4   & 2   & 6   & 8       & 1  &       X       &      -10.0609     &          -2.16831     &    0.00        \\
4   & 2   & 6   & 8       & 2  &    2.45356    &      -2.58269     &           1.09179     &     180.00     \\
4   & 2   & 6   & 8       & 3  &        X      &        -1.46073   &          0.192876     &    0.00        \\
4   & 2   & 6   & 11      & 1  &        X      &          X        &          -2.89358     &    0.00        \\
4   & 2   & 6   & 11      & 2  &        X      &          X        &           1.75806     &    180.00      \\
4   & 2   & 6   & 11      & 3  &        X      &          X        &           0.230563    &     0.00       \\
5   & 3   & 7   & 6       & 1  &        X      &        -9.15296   &          -2.89358     &    0.00        \\
5   & 3   & 7   & 6       & 2  &    -0.999624  &       -1.25200    &           1.75806     &    180.00      \\
5   & 3   & 7   & 6       & 3  &       X       &        -0.997410  &          0.230563     &    0.00        \\
5   & 3   & 7   & 9       & 1  &       X       &        -10.0609   &          -2.16831     &    0.00        \\
5   & 3   & 7   & 9       & 2  &    2.45356    &       -2.58269    &           1.09179     &     180.00     \\
5   & 3   & 7   & 9       & 3  &       X       &      -1.46073     &           0.192876    &     0.00       \\
5   & 3   & 7   & 12      & 1  &       X       &      -9.15296     &          -2.89358     &    0.00        \\
5   & 3   & 7   & 12      & 2  &       X       &      -1.25200     &           1.75806     &    180.00      \\
5   & 3   & 7   & 12      & 3  &       X       &      -0.997410    &           0.230563    &     0.00       \\
6   & 7   & 12  &  10     & 1  &       X       &      -6.25427     &          -1.53021     &    0.00        \\
6   & 7   & 12  &  10     & 2  &       X       &      -0.819813    &          -0.713956    &      180.00    \\
6   & 7   & 12  &  10     & 3  &       X       &      -1.40487     &          -0.0941236   &      0.00   \\   
6   & 7   & 12  &  16     & 1  &       X       &      0.0503902    &          -1.74289     &     0.00       \\
6   & 7   & 12  &  16     & 2  &       X       &      1.43493      &         -2.69254     &    180.00      \\ 
6   & 7   & 12  &  16     & 3  &       X       &      5.08310      &          -2.56013     &    0.00        \\
6   & 7   & 12  &  17     & 1  &       X       &      -6.25427     &          -1.53021     &    0.00        \\
6   & 7   & 12  &  17     & 2  &       X       &      -0.819813    &          -0.713956    &      180.00    \\
6   & 7   & 12  &  17     & 3  &       X       &      -1.40487     &          -0.0941236   &     0.00    \\   
6   & 11  &  14 &   16    & 1  &       X       &       X           &        -2.46899     &     0.00        \\ 
6   & 11  &  14 &   16    & 2  &       X       &        X          &         0.0047732   &   180.00     \\    
6   & 11  &  14 &   16    & 3  &       X       &        X          &        -0.713956    &      0.00       \\ 
6   & 11  &  14 &   18    & 1  &       X       &        X          &         0.000       &      0.00       \\ 
7   & 6   & 11  &  13     & 1  &       X       &        X          &        -1.53021     &    0.00         \\ 
7   & 6   & 11  &  13     & 2  &       X       &        X          &        -0.713956    &     180.00      \\ 
7   & 6   & 11  &  13     & 3  &       X       &        X          &        -0.094123    &    0.00        \\  
7   & 6   & 11  &  14     & 1  &       X       &        X          &        -1.74289     &     0.00        \\ 
7   & 6   & 11  &  14     & 2  &       X       &        X          &        -2.69254     &    180.00       \\ 
7   & 6   & 11  &  14     & 3  &       X       &        X          &        -2.56013     &    0.00         \\ 
7   & 6   & 11  &  15     & 1  &       X       &        X          &        -1.53021     &    0.00         \\ 
7   & 6   & 11  &  15     & 2  &       X       &        X          &        -0.713956    &     180.00      \\ 
7   & 6   & 11  &  15     & 3  &       X       &        X          &        -0.0941236   &     0.00         \\
7   & 12  &  16 &   14    & 1  &       X       &      1.56618      &        -1.74289     &     0.00        \\ 
7   & 12  &  16 & 14    & 2  &       X       &      -2.45376     &        0.0047732   &    180.00      \\   
7   & 12  &  16 & 14    & 3  &       X       &      -0.819813    &       -0.713956    &      0.00       \\   
7   & 12  &  16 & 19    & 1  &       X       &      0.000        &        0.000       &     0.00        \\   
8   & 6   & 7  &  9       & 1  &       X       &      -3.43664     &       -0.0510087   &     0.00       \\   
8   & 6   & 7  &  9       & 2  &    2.34735    &      1.24857      &       -2.28477     &     180.00      \\  
8   & 6   & 7  &  9       & 3  &       X       &       1.78467     &       0.0546769    &    0.00         \\  
8   & 6   & 7  &  12      & 1  &       X       &       -6.25427    &      -1.53021      &    0.00          \\ 
8   & 6   & 7  &  12      & 2  &       X       &       -0.819813   &      -0.713956     &     180.00      \\  
8   & 6   & 7  &  12      & 3  &       X       &       -1.40487    &      -0.094123     &     0.00     \\     
8   & 6   & 11 &   13     & 1  &       X       &         X         &      -0.051008     &    0.00     \\      
8   & 6   & 11 &   13     & 2  &       X       &         X         &      -2.28477      &    180.00       \\ 
   8   & 6   & 11 &   13     & 3  &       X       &         X         &       0.0546769    &    0.00     \\      
   8   & 6   & 11 &   14     & 1  &       X       &         X         &        0.861788     &    0.00         \\ 
   8   & 6   & 11 &   14     & 2  &       X       &         X         &        0.995909     &     180.00      \\ 
   8   & 6   & 11 &   14     & 3  &       X       &         X         &        -1.07686     &    0.00         \\ 
   8   & 6   & 11 &   15     & 1  &       X       &         X         &        -0.051008    &     0.00    \\     
   8   & 6   & 11 &   15     & 2  &       X       &         X         &        -2.28477     &     180.00      \\ 
   8   & 6   & 11 &   15     & 3  &       X       &         X         &        0.0546769    &     0.00     \\    
   9   & 7   & 6  &  11      & 1  &       X       &         X         &        -1.53021     &    0.00         \\ 
   9   & 7   & 6  &  11      & 2  &       X       &         X         &        -0.713956    &      180.00     \\ 
   9   & 7   & 6  &  11      & 3  &       X       &         X         &        -0.0941236   &      0.00    \\    
   9   & 7   & 12 &   10     & 1  &       X       &       -3.43664    &       -0.0510087    &     0.00     \\    
   9   & 7   & 12 &   10     & 2  &       X       &       1.24857     &       -2.28477      &    180.00       \\ 
   9   & 7   & 12 &   10     & 3  &       X       &       1.78467     &        0.054676     &    0.00      \\    
   9   & 7   & 12 &   16     & 1  &       X       &       -1.19475    &        0.861788     &    0.00          \\
   9   & 7   & 12 &   16     & 2  &       X       &       3.74789     &        0.995909     &     180.00       \\
   9   & 7   & 12 &   16     & 3  &       X       &       -1.42060    &        -1.07686     &    0.00          \\
   9   & 7   & 12 &   17     & 1  &       X       &       -3.43664    &       -0.0510087    &     0.00      \\   
   9   & 7   & 12 &   17     & 2  &       X       &       1.24857     &        -2.28477     &     180.00       \\
   9   & 7   & 12 &   17     & 3  &       X       &       1.78467     &        0.0546769    &     0.00      \\   
   10  &  12 &  16  &  14   & 1  &       X       &       -4.27768    &        -3.29003     &    0.00          \\ 
   10  &  12 &  16  &  14   & 2  &    3.21391    &        3.74789    &         0.995909    &      180.00      \\ 
   10  &  12 &  16  &  14   & 3  &       X       &      0.960659     &        -0.693868    &      0.00        \\ 
   10  &  12 &  16  &  19   & 1  &       X       &      2.13188      &           0.000     &     0.00         \\ 
   10  &  12 &  16  &  19   & 2  &    2.53826    &     -3.11705      &             X       &        X         \\
   11  &  6  &  7   & 12     & 1  &       X       &         X         &        -0.601273    &     0.00         \\
   11  &  6  &  7   & 12     & 2  &       X       &         X         &         5.54093     &    180.00        \\
   11  &  6  &  7   & 12     & 3  &       X       &         X         &        -1.18306     &    0.00          \\
   11  &  14 &  16  &  12   & 1  &    -0.818383  &      -2.27656     &        -5.28649     &     0.00         \\ 
   11  &  14 &  16  &  12   & 2  &    1.645260   &      3.43143      &         2.36666     &    180.00        \\ 
   11  &  14 &  16  &  12   & 3  &    -1.23972   &      -2.35012     &            X        &       X          \\
   11  &  14 &  16  &  19   & 2  &    3.07771    &      0.554677     &         2.36666     &    180.00        \\ 
   12  &  16 &  14  &  18   & 2  &    3.07771    &      0.554677     &         2.36666     &    180.00        \\ 
   13  &  11 &  14  &  16   & 1  &       X       &      -4.27768     &        -3.29003     &    0.00          \\ 
   13  &  11 &  14  &  16   & 2  &    2.42160    &      3.74789      &        0.995909     &   180.00        \\  
   13  &  11 &  14  &  16   & 3  &       X       &      0.960659     &        -0.693868    &    0.00          \\
   13  &  11 &  14  &  18   & 1  &       X       &      0.000        &        0.000      &    0.00          \\ 
   13  &  11 &  14  &  18   & 2  &    2.53826    &    -3.11705      &           X       &      X           \\
   14  &  16 &  12  &  17   & 1  &       X       &   -4.27768      &       -3.29003    &     0.00         \\  
   14  &  16 &  12  &  17   & 2  &    3.21391    &    3.74789       &       0.995909    &      180.00      \\  
   14  &  16 &  12  &  17   & 3  &       X       &    0.960659      &      -0.693868    &      0.00        \\  
   15  &  11 &  14  &  16   & 1  &       X       &   -4.27768       &       -3.29003    &     0.00         \\  
   15  &  11 &  14  &  16   & 2  &    2.42160    &   3.74789        &       0.995909    &      180.00      \\  
   15  &  11 &  14  &  16   & 3  &       X       &   0.960659       &      -0.693868    &     0.00         \\ 
   15  &  11 &  14  &  18   & 1  &       X       &    0.000         &        0.000      &       0.00       \\    
   15  &  11 &  14  &  18   & 2  &    2.53826    &   -3.11705       &           X       &        X         \\
   17  &  12 &  16  &  19   & 1  &       X       &     0.000       &         0.000     &        0.00      \\     
   17  &  12 &  16  & 19   & 2  &    2.53826    &    -3.11705      &          X        &       X          \\
   18  &  14 &  16  & 19   & 2  &    3.13276    &    3.43143       &        2.36666    &     180.00       \\

\label{tabS4}
    \end{longtable}  

   \begin{table}[H] 
   \scriptsize
   \caption{Non-bonded parameters of the MS-ARMD reactant PES. ``X'' indicates that this parameter is not needed.}
   \resizebox{\textwidth}{!}{
   \begin{tabular}{c|c|c|c|c|c} 
   Atom \# & $q_i$ [e] & $\epsilon_{i,1}$ [kcal/mol]& $R_{min,1}$/2[\AA] &$\epsilon_{i,2}$ [kcal/mol]& $R_{min,2}$/2[\AA] \\ \hline  
	1   &  -0.300000   &  0.203207E-04  &  2.80541   &  X    & X      \\
	2   &   0.705600   &  0.667655      &  0.102478  &  X    & X      \\
	3   &   0.705600   &  0.667655      &  0.102478  &  X    & X      \\
	4   &  -0.570000   &  0.411920E-08  &  4.77007   & 0.120 &  1.40  \\
	5   &  -0.570000   &  0.411920E-08  &  4.77007   & 0.120 &  1.40  \\
	6   &  -0.135600   &  0.242868E-01  &  2.13203   &  X    & X      \\
	7   &  -0.135600   &  0.242868E-01  &  2.13203   &  X    & X      \\
	8   &   0.150000   &  0.218559      &  0.917005  &  X    & X      \\
	9   &   0.150000   &  0.218559      &  0.917005  &  X    & X      \\
	10  &   0.152910   &  0.218559      &  0.917005  &  X    & X      \\
	11  &   0.221270   &  0.242868E-01  &  2.13203   &  X    & X      \\
	12  &   0.221270   &  0.242868E-01  &  2.13203   &  X    & X      \\
	13  &   0.152910   &  0.218559      &  0.917005  &  X    & X      \\
	14  &  -0.151030   &  0.242868E-01  &  2.13203   &  X    & X      \\
	15  &   0.152910   &  0.218559      &  0.917005  &  X    & X      \\
	16  &  -0.151030   &  0.242868E-01  &  2.13203   &  X    & X      \\
	17  &   0.152910   &  0.218559      &  0.917005  &  X    & X      \\
	18  &   0.139550   &  0.648973E-01  &  2.32950   &  X    & X      \\
	19  &   0.139550   &  0.648973E-01  &  2.32950   &  X    & X      \\
	
\hline

Atom 1 \# & Atom 2\# & $\epsilon_{i}$ [kcal/mol] & $R_{min}$/2[\AA] & $n$ & $m$ \\ \hline
18    &  4   &  2.76499  & 3.30525  & 15.5903  & 16.2486     \\
18    &  5   &  2.76499  & 3.30525  & 15.5903  & 16.2486     \\
19    &  4   &  2.76499  & 3.30525  & 15.5903  & 16.2486     \\
19    &  5   &  2.76499  & 3.30525  & 15.5903  & 16.2486     \\
11    &  6   &  6.23430  & 2.25895  & 3.34975  & 5.24917     \\
12    &  7   &  6.23430  & 2.25895  & 3.34975  & 5.24917     \\
11    &  7   &  6.23430  & 2.25895  & 3.34975  & 5.24917     \\
12    &  6   &  6.23430  & 2.25895  & 3.34975  & 5.24917     \\

   \end{tabular} }
   \label{tabS5}
   \end{table}   
   
   \begin{table}[H] 
      \scriptsize
   \caption{Non-bonded parameters of the MS-ARMD intermediate
     PES. ``X'' indicates that this parameter is not needed.}
   \resizebox{\textwidth}{!}{
   \begin{tabular}{c|c|c|c|c|c} 
   Atom \# & $q_i$ [e] & $\epsilon_{i,1}$ [kcal/mol]& $R_{min,1}$/2[\AA] &$\epsilon_{i,2}$ [kcal/mol]& $R_{min,2}$/2[\AA] \\ \hline  
   1   &  -0.287170       & 0.152100  &  1.770000   & X     & X       \\
   2   &   0.545440       & 0.110000  &  2.000000   & X     & X       \\
   3   &   0.545440       & 0.110000  &  2.000000   & X     & X       \\
   4   &  -0.392740       & 0.120000  &  1.700000   & 0.120 &  1.40   \\
   5   &  -0.392740       & 0.120000  &  1.700000   & 0.120 &  1.40   \\
   6   &  -0.408260E-01   & 0.055000  &  2.175000   & 0.010 &  1.90   \\
   7   &  -0.408260E-01   & 0.055000  &  2.175000   & 0.010 &  1.90   \\
   8   &   0.104730       & 0.022000  &  1.320000   & X     & X       \\
   9   &   0.104730       & 0.022000  &  1.320000   & X     & X       \\
   10  &   0.195520       & 0.022000  &  1.320000   & X     & X       \\
   11  &  -0.803190E-01   & 0.055000  &  2.175000   & 0.010 &  1.90   \\
   12  &  -0.803190E-01   & 0.055000  &  2.175000   & 0.010 &  1.90   \\
   13  &   0.195520       & 0.022000  &  1.320000   & X     & X       \\
   14  &  -0.914950E-01   & 0.068000  &  2.090000   & X     & X       \\
   15  &   0.883190E-01   & 0.022000  &  1.320000   & X     & X       \\
   16  &  -0.914950E-01   & 0.068000  &  2.090000   & X     & X       \\
   17  &   0.883190E-01   & 0.022000  &  1.320000   & X     & X       \\
   18  &   0.321120       & 4.35177   &  1.69104    & X     & X       \\
   19  &   0.321120       & 4.35177   &  1.69104    & X     & X       \\
   
\hline

Atom 1 \# & Atom 2\# & $\epsilon_{i}$ [kcal/mol] & $R_{min}$/2[\AA] & $n$ & $m$ \\ \hline
  18    &  4   &  5.24581  & 3.55543  & 3.04411  & 4.51621  \\ 
  18    &  5   &  5.24581  & 3.55543  & 3.04411  & 4.51621  \\ 
  19    &  4   &  5.24581  & 3.55543  & 3.04411  & 4.51621  \\ 
  19    &  5   &  5.24581  & 3.55543  & 3.04411  & 4.51621  \\ 
  11    &  6   &  4.20015  & 1.90058  & 3.53818  & 5.54809  \\ 
  11    &  7   &  4.20015  & 1.90058  & 3.53818  & 5.54809  \\

   \end{tabular} }
   \label{tabS6}
   \end{table}   
      
   \begin{table}[H] 
      \scriptsize
   \caption{Non-bonded parameters of the MS-ARMD product PES. ``X''
     indicates that this parameter is not needed.}
   \resizebox{\textwidth}{!}{
   \begin{tabular}{c|c|c|c|c|c} 
   Atom \# & $q_i$ [e] & $\epsilon_{i,1}$ [kcal/mol]& $R_{min,1}$/2[\AA] &$\epsilon_{i,2}$ [kcal/mol]& $R_{min,2}$/2[\AA] \\ \hline 
	  1   &  -0.285000      & 0.152100   & 1.770000  & X      & X      \\
	  2   &   0.568000      & 0.110000   & 2.000000  & X      & X      \\
	  3   &   0.568000      & 0.110000   & 2.000000  & X      & X      \\
	  4   &  -0.414000      & 0.120000   & 1.700000  & 0.120  & 1.40   \\
	  5   &  -0.414000      & 0.120000   & 1.700000  & 0.120  & 1.40   \\
	  6   &  -0.034000  & 0.055000   & 2.175000  & 0.010  & 1.90   \\
	  7   &  -0.034000  & 0.055000   & 2.175000  & 0.010  & 1.90   \\
	  8   &   0.11240       & 0.022000   & 1.320000  & X      & X      \\
	  9   &   0.11240       & 0.022000   & 1.320000  & X      & X      \\
	  10  &   0.19260       & 0.022000   & 1.320000  & X      & X      \\
	  11  &  -0.055200      & 0.055000   & 2.175000  & 0.010  & 1.90   \\
	  12  &  -0.055200      & 0.055000   & 2.175000  & 0.010  & 1.90   \\
	  13  &   0.19260       & 0.022000   & 1.320000  & X      & X      \\
	  14  &  -0.126000      & 0.068000   & 2.090000  & X      & X      \\
	  15  &   0.08290       & 0.022000   & 1.320000  & X      & X      \\
	  16  &  -0.126000      & 0.068000   & 2.090000  & X      & X      \\
	  17  &   0.08290       & 0.022000   & 1.320000  & X      & X      \\
	  18  &   0.317000      & 3.99998    &  1.67206  & X      & X      \\
	  19  &   0.317000      & 3.99998    &  1.67206  & X      & X      \\
	    
   \end{tabular} }
   \label{tabS7}
   \end{table}    

\noindent
The barrier region connecting the reactant and intermediate force
fields and the intermediate and product force fields is described by
two GAPOs $ \Delta V^{ij}_{\rm GAPO,k} (x) = exp\left( -\frac{(\Delta
  V_{ij}(x)-V^0_{ij,k})^2}{2 \sigma^2_{ij,k}}\right) \times
\sum_{l=0}^{m_{ij,k}} a_{ij,kl} (\Delta V_{ij}(x)-V^0_{ij,k})^l$ with
the parameters summarized in Table \ref{tabS8}.

\begin{table}[H]
\caption{GAPO parameters: $i$ labels the reactant (R$^+$) or the
  intermediate (INT$^+$) and $j$ labels the intermediate or the
  product (P$^+$), $V^0_{ij,k}$ is the center of the Gaussian function
  (in kcal/mol), and $\sigma_{ij,k}$ the width of the Gaussian (in
  kcal/mol). $a_{ij}$ is the polynomial coefficient in (kcal/mol)$^{1
    - j}$, $j = 0, 3$.}  \resizebox{\textwidth}{!}{
\begin{tabular}{c|c|c|c|c|c|c|c|c} 
  $i$ & $j$ & $k$ & $V^0_{ij,k}$ & $\sigma_{ij,k}$ & $a_{ij,k0}$ & $a_{ij,k1}$ & $a_{ij,k2}$ & $a_{ij,k3}$\\
  \hline
     R$^+$ & INT$^+$ & 3 & 2.2385E+01 & 2.5180E+01 &-1.5000E+01 & 3.9005E-01 & -1.0249E-02 & 1.1676-04 \\
     INT$^+$ & P$^+$&  2 & -2.8049E+01 & 3.5355E+01 &-1.2000E+01 & -4.8562E-01 & -5.6894E-03 & {} \\
   \end{tabular} }
\label{tabS8}
\end{table}